\documentclass[preprint,12pt,authoryear]{elsarticle}

\usepackage{graphicx}
\usepackage{amsmath}
\usepackage{amssymb}
\usepackage{amsthm}
\usepackage{derivative}

\usepackage{lineno}

\usepackage{subfigure}

\usepackage{caption}
\usepackage{newtxtext}
\usepackage{newtxmath}

\usepackage{float}

\usepackage{hyperref}
\hypersetup{
    colorlinks = true,
    urlcolor   = black,
    citecolor  = black,
}

\usepackage[svgnames]{xcolor}
\usepackage{ulem}

\usepackage{multicol}
\usepackage{multirow}
\usepackage{adjustbox}

\usepackage{svg}

\usepackage{tikz}

\usepackage[ruled, linesnumbered]{algorithm2e}
\SetKwInput{KwInput}{Input}                
\SetKwInput{KwOutput}{Output}              
\usepackage{setspace} 

\usepackage{amssymb}
\usepackage{nomencl}
\makenomenclature

\usepackage{etoolbox}
\renewcommand\nomgroup[1]{%
  \item[\bfseries
  \ifstrequal{#1}{A}{General symbols}{%
  \ifstrequal{#1}{B}{Greek symbols}{%
  \ifstrequal{#1}{C}{Abbreviations}{
  \ifstrequal{#1}{D}{Superscripts/subscripts}{}}}}%
]}

\journal{Int. J. Heat and Fluid Flow}

\begin{document}


\begin{frontmatter}

\title{Characterization of supersonic boundary layers \\ of adiabatic and isothermal curved surfaces \\ with shock interactions}

\author[inst1,inst2]{Gabriel Y. R. Hamada}
\author[inst1]{William R. Wolf} 
\author[inst1]{Hugo F. S. Lui}
\author[inst2]{Carlos Junqueira-Junior}

\affiliation[inst1]{organization={Faculdade de Engenharia Mecânica, Universidade Estadual de Campinas},
            city={Campinas},
            postcode={13086-860},
            country={Brazil}}
\affiliation[inst2]{organization={DynFluid, Arts et Métiers Institute of Technology, CNAM, 151 Boulevard de l’Hôpital},
	city={Paris},
	postcode={75013},
	country={France}}

\begin{abstract}
Boundary layers of adiabatic and isothermal curved walls are investigated for a supersonic turbine cascade, including the effects of shock-boundary layer interactions (SBLIs). Wall-resolved large eddy simulations (LES) are performed for a linear cascade of blades with an inlet Mach number of $M_\infty = 2.0$ and Reynolds number based on the axial chord $Re_\infty = 200\,000$. The wall to inlet temperature ratio of the isothermal case is $T_w/T_{\infty}=0.75$, representing a cooled wall. An assessment of the effects of pressure gradient, thermal boundary conditions and SBLIs is presented in terms of the downstream variation of mean flow quantities such as density, temperature, and momentum profiles. The different thermal boundary conditions affect the density and temperature profiles along the boundary layer, where cooling increases the density of the gas near the wall, and reduces its temperature and viscosity. Both of these effects make the momentum profiles fuller and, hence, the boundary layer of the isothermal case is less prone to separate than that of the adiabatic wall. The mean density profiles are also affected by pressure gradients induced by the convex and concave curvatures of the blade, which lead to expansion and compression of the flow, respectively. The analysis of separate terms from the momentum balance equation explains the behavior of various physical mechanisms in the inner and outer regions of the supersonic boundary layers. The importance of mean flow advection, compressibility, and Reynolds stresses is presented in terms of flow acceleration and deceleration. The impact of the SBLIs in the momentum balance mechanisms is also investigated, showing that a combination of compressions and expansions impact the boundary layers by redirecting the flow toward the wall due to the shock formations.
\end{abstract}

\begin{keyword}
Supersonic boundary layer \sep pressure gradient \sep shock-boundary layer interaction \sep adiabatic wall \sep isothermal wall
\end{keyword}

\end{frontmatter}

\nomenclature[A]{$c_x$}{Blade axial chord length}
\nomenclature[A]{$M_\infty$}{Inlet Mach number}
\nomenclature[A]{$Re_\infty$}{Chord-based Reynolds number}
\nomenclature[A]{$Pr$}{Prandtl number}
\nomenclature[A]{$H$}{Shape factor}

\nomenclature[B]{$\beta$}{Clauser pressure gradient parameter}
\nomenclature[B]{$\delta$}{Boundary layer thickness}
\nomenclature[B]{$\delta^*$}{Boundary layer displacement thickness}
\nomenclature[B]{$\theta$}{Boundary layer momentum thickness}
\nomenclature[B]{$\kappa$}{Local surface curvature}

\nomenclature[C]{SBLI}{Shock-boundary layer interaction}
\nomenclature[C]{FPG}{Favorable pressure gradient}
\nomenclature[C]{APG}{Adverse pressure gradient}
\nomenclature[C]{DNS}{Direct numerical simulation}
\nomenclature[C]{LES}{Large eddy simulation}
\nomenclature[C]{PS}{Pressure side}
\nomenclature[C]{SS}{Suction side}
\nomenclature[C]{S-U-F}{Suction side - Upstream of the SBLI - Favorable pressure gradient}
\nomenclature[C]{S-D-F}{Suction side - Downstream of the SBLI - Favorable pressure gradient}
\nomenclature[C]{P-U-A}{Pressure side -  Upstream of the SBLI - Adverse pressure gradient}
\nomenclature[C]{P-D-F}{Pressure side -  Downstream of the SBLI - Favorable pressure gradient}

\nomenclature[D]{$\infty$}{Property evaluated at the cascade inlet}
\nomenclature[D]{$e$}{Property evaluated at the boundary layer edge}
\nomenclature[D]{$w$}{Property evaluated at the wall}
\nomenclature[D]{$n$}{Property evaluated in the wall-normal direction}
\nomenclature[D]{$+$}{Property evaluated in wall units}
\printnomenclature

\section{Introduction}

Supersonic boundary layers are present in engineering applications such as high-speed vehicles \citep{liu2023experimental, bhagwandin2023wall} and turbomachinery \citep{liu2019characterization}. These devices may present curved surfaces for lift production or other aerodynamic requirements such as in control surfaces of supersonic aircraft and supersonic turbine blades. The latter are coupled to detonation engines and, for these cases, have been shown by \citet{paniagua2014design} to be more efficient than conventional subsonic turbines. Despite their previous applications in aeronautical and mechanical engineering systems, the study of supersonic boundary layers developing over curved surfaces is not widely covered in the literature. \citet{lui2024mach} shows that the streamwise surface curvature of supersonic turbines leads to changes in the state of the boundary layers that directly impact the shock-boundary layer interactions (SBLIs). Variations in the state of compressible boundary layers can occur due to surface curvature, and its induced pressure gradients, besides thermal boundary conditions. Understanding the impact of such parameters in the development of supersonic boundary layers is important for designing more efficient power generation systems and high-speed transportation.

The impact of surface thermal conditions has been reported for supersonic boundary layers \citep{coleman1995numerical,morinishi,trettel2016mean,schiavo2021high,hirai2021effects,cogo2022direct,cogo2023assessment} and SBLIs \citep{bernardini2016heat,volpiani2018effects,volpiani2020effects}. \citet{cogo2022direct, cogo2023assessment} performed direct numerical simulations (DNS) of zero pressure gradient turbulent boundary layers on flat plates with different thermal boundary conditions.  Adiabatic and cooled walls were studied for Mach numbers ranging from $2$ to $6$ and it was observed that wall cooling induces a temperature peak along the boundary layer, leading to a stratification of flow properties on the buffer layer. This stratification enhances the velocity fluctuations in the streamwise direction while the other velocity components are damped. 

\citet{bernardini2016heat} investigated the impact of thermal boundary conditions in SBLIs using DNS of zero pressure gradient turbulent boundary layers. Simulations were conducted for Mach number $2.28$ where an oblique shock impinges on a flat plate. Different wall-to-recovery temperature ratios were analyzed including adiabatic, cooled and heated walls. This work was later extended by \citet{volpiani2018effects, volpiani2020effects}, who observed that the magnitudes of the flow properties along the SBLI are significantly affected by the wall temperature, with wall cooling leading to a reduction of the separation length and interaction scales when compared to adiabatic and heated cases. At the same time, wall cooling also led to higher dynamic loads on the surface, which can be structurally detrimental. 

The impact of variable pressure gradients in supersonic boundary layers was discussed by \cite{spina1994physics}, who compiled experimental results obtained by several authors. The opposite effects of adverse and favorable pressure gradients (APGs and FPGs) in the streamwise direction were discussed for turbulent boundary layers, with the adverse pressure gradient leading to a compression of streamtubes while the favorable pressure gradient lead to its dilatation. Hence the APG can lead to an increase of the turbulence intensity while the FPG, depending on its strength, can lead to flow relaminarization. The work also mentioned that adverse pressure gradients lead to thinner boundary layers while the opposite behavior is observed for favorable pressure gradients. \citet{wenzel2022influences} employed DNS to investigate heat transfer and pressure gradient effects on turbulent boundary layers. They evaluated momentum and energy transfers using integral identities and showed that the dependence of Mach number and wall temperature are accounted for by the Eckert number. More recently, \citet{wen2023response} studied the impact of pressure gradient through experiments conducted in a supersonic wind tunnel. Turbulent boundary layers under adverse pressure gradients were investigated for Mach $2.7$. The authors corroborated that an increase in the APG leads to a thinner boundary layer, and it was also shown that the Reynolds shear stresses and turbulent production are enhanced. 

Alongside the state of the boundary layer, the shock strength also impacts the SBLI as shown by \citet{morgan2013flow}, who performed large eddy simulations (LES) of flows with different shock strengths and Reynolds numbers. It was shown that small variations in shock strength could change the SBLI from an incipient regime to a fully separated one. \citet{volpiani2020effects} observed the impact of shock strength in a zero pressure gradient turbulent boundary layer under hypersonic flow conditions. They observed that the skin friction coefficient inside the separation region displays a plateau before reaching a peak and reattaching when the shock strength is strong. 

Most of the aforementioned studies were conducted for canonical flow configurations, such as plane channels and flat plates. Moreover, the boundary layers were fully developed before evaluating the impacts of thermal boundary conditions and pressure gradients on the SBLIs. In a more realistic flow setup, such as a supersonic turbine, the boundary layer may not have time to fully develop along the blade. Hence, the SBLI can occur under a transitional regime. The impact of the boundary layer regime on SBLIs was investigated by \citet{sandham2014transitional}, who utilized experiments and numerical simulations of transitional flows with zero pressure gradient at Mach $6$. They showed that higher levels of wall heat transfer were obtained for transitional boundary layers rather than fully turbulent ones. As observed by \citet{hamada2024characterization} for a supersonic turbine cascade, the transitional regime is more likely to occur on the suction side, where the boundary layer is subjected to a favorable pressure gradient  which tends to relaminarize \citep{spina1994physics}.

A considerable difference between a supersonic turbine and a zero pressure gradient configuration is the presence of variable streamwise pressure gradients induced by the blade curvature. While the suction side has a convex curvature which leads to a favorable pressure gradient, the pressure side is concave and the boundary layer is subject to an adverse pressure gradient. The SBLIs on both sides of the blade are also impacted by the pressure gradient. Considering the supersonic turbine cascade studied by \citet{liu2019characterization,lui2022unsteadiness}, an oblique shock impinges on the suction side boundary layer, whereas a Mach reflection forms on the pressure side. For the same turbine cascade, \citet{lui2024mach} studied the effects of inlet Mach number variations in the SBLIs. Their results indicated that an increase in the Mach number leads to a weaker impinging shock on the suction side but a larger separation bubble. The opposite trend was observed on the pressure side, but the separation bubble decreased only slightly, showing that flow separation in SBLIs depends not only on the shock strength, but also on the state of the incoming boundary layer, curvature and pressure gradient.

To the authors' knowledge, \citet{namatsaliuk2025experimental} performed the first experiment for the supersonic turbine configuration studied by \citet{lui2022unsteadiness}. In the experiments, they found a similar flow topology on the pressure side of the turbine. However, some differences were observed downstream of the SBLI on the suction side, which depicts a larger separation compared to the LES from the previous authors. It should be mentioned that the experiments were conducted for a higher Reynolds number than the LES, and without boundary layer tripping. Some small discrepancies in the blade angle of attack were also reported in the experiment. \citet{hamada2023thermal} investigated the impact of adiabatic and isothermal (cooled) boundary conditions of the flow over the previous supersonic turbine cascade. Their work remarks that a cooled wall leads to higher levels of turbulent kinetic energy production and dissipation on the suction side, upstream of the SBLI. Two production peaks were shown after the SBLI, indicating the presence of a shear layer embedded in the turbulent boundary layer \citep{leandro2024}. In accordance with other studies \citep{bernardini2016heat,volpiani2018effects,volpiani2020effects}, wall cooling leads to smaller separation regions due to a local reduction in viscosity that increases the Reynolds number.  

Due to the finite blade chord, the boundary layer of a supersonic turbine may be laminar, transitional or turbulent depending on the inlet flow features and the presence of tripping. Moreover, the blade curvature induces a streamwise pressure gradient and surface heat transfer is impacted by thermal boundary conditions. In this work, we investigate the impact of adverse and favorable pressure gradients on boundary layers of adiabatic and isothermal (cooled) blades of supersonic turbine cascades. A study is presented in terms of thermodynamic and kinematic mean flow quantities and includes a characterization of the supersonic boundary layer states upstream and downstream of SBLIs. Results are obtained from wall-resolved  LES of a supersonic turbine cascade with inlet Reynolds and Mach numbers $Re_\infty = 200,000$ and $M_\infty = 2.0$, respectively. The present work is structured as follows: the theoretical and numerical formulations, as well as the flow configuration, are presented in \S \ref{sec:theoreticalformulation}. Results follow in \S \ref{sec:results}, where the flow characterization is presented together with a discussion on the streamwise variation of thermodynamic and kinematic properties in the boundary layers. The main conclusions of this study are presented in \S \ref{sec:conclusions}.

\section{Theoretical formulation and numerical methods}\label{sec:theoreticalformulation}

\subsection{Governing equations}

The compressible Navier-Stokes equations are solved in a generalized curvilinear coordinate system ($\xi^i, i = 1,2,3$) with the velocity components written in the contravariant form ($u^i$). As shown by \citet{aris1990vectors}, the mass, momentum, and energy balance equations can be expressed as
\begin{equation}\label{eq:continuity_equation}
\pdv{\rho}{t} +(\rho u^j)_{,j} = 0,
\end{equation}
\begin{equation}\label{eq:momentum_equation}
\pdv{\rho u^i}{t} +\left(\rho u^iu^j + g^{ij}p - \tau^{ij}\right)_{,j} = f^{i}_{b},
\end{equation}
and
\begin{equation}\label{eq:energy_equation}
\pdv{E}{t} +\left[(E+p)u^j +q^j -\tau^{ij}g_{ik}u^k\right]_{,j} =  g_{ij}u^jf^i_b.
\end{equation}
Here, the time is represented by $t$, density by $\rho$, pressure by $p$. The terms $g_{ij}$ and $g^{ij}$ are the covariant and contravariant metric tensors, respectively. Subscripts preceded by a comma ${(\cdot)_{,j}}$ represent covariant derivatives (see \ref{sec:appendixATensor} for details), and $f^i_b$ is a contravariant body force component. The Einstein summation convention applies for repeated indices.

Considering a Newtonian fluid obeying Fourier's law, the viscous stress tensor ($\tau^{ij}$), heat flux ($q^j$), and total energy ($E$) are given, respectively, by
\begin{equation}\label{eq:viscous_stress_tensor}
\tau^{ij} = \mu\dfrac{M_\infty}{Re_{\infty}}\left(g^{jk}u^i_{,k} 
          + g^{ik}u^j_{,k} -\dfrac{2}{3}g^{ij}u^k_{,k}\right) \,\mbox{,}
\end{equation}
\begin{equation}\label{eq:heat_flux}
q^j = -\dfrac{M_\infty}{Re_{\infty}Pr}\mu g^{ij}T_{,i}\,\mbox{,}
\end{equation}
and
\begin{equation}\label{eq:total_energy}
E = \dfrac{p}{\gamma-1}+\dfrac{1}{2}\rho g_{ij}u^iu^j\,\mbox{,}
\end{equation}
where $\gamma$ is the ratio of specific heats, and $T$ is the temperature. Assuming a calorically perfect gas, the set of equations is closed by the equation of state
\begin{equation}\label{eq:equation_of_state}
p = \dfrac{(\gamma-1)}{\gamma}\rho T\,\mbox{,}
\end{equation}
and the dynamic viscosity $\mu$ is computed by Sutherland's law, written as
\begin{equation}\label{eq:sutherlandslaw}
\mu = \left[(\gamma-1)T\right]^{\frac{3}{2}}\dfrac{1+S_\mu}{T(\gamma-1)+S_\mu}\,\mbox{,}
\end{equation}
where $S_\mu$ is the non-dimensional Sutherland constant.

The governing equations are written in non-dimensional form where $Re_{\infty}$, $M_\infty$ and $Pr$ are the Reynolds, Mach and Prandtl numbers, respectively. The subscript $\infty$ refers to properties computed at the cascade inlet, and all flow parameters and transport coefficients are non-dimensionalized by inlet properties and the blade axial chord length $(c_x)$. Further details regarding the non-dimensionalization procedure are provided by \citet{lui2022unsteadiness}.

\subsection{Flow configuration, grid setup and numerical schemes}

\begin{figure}[H]
	\centering
	\includegraphics[trim={0.25cm 1.2cm 0.7cm 3.05cm},clip,width=0.75\textwidth]{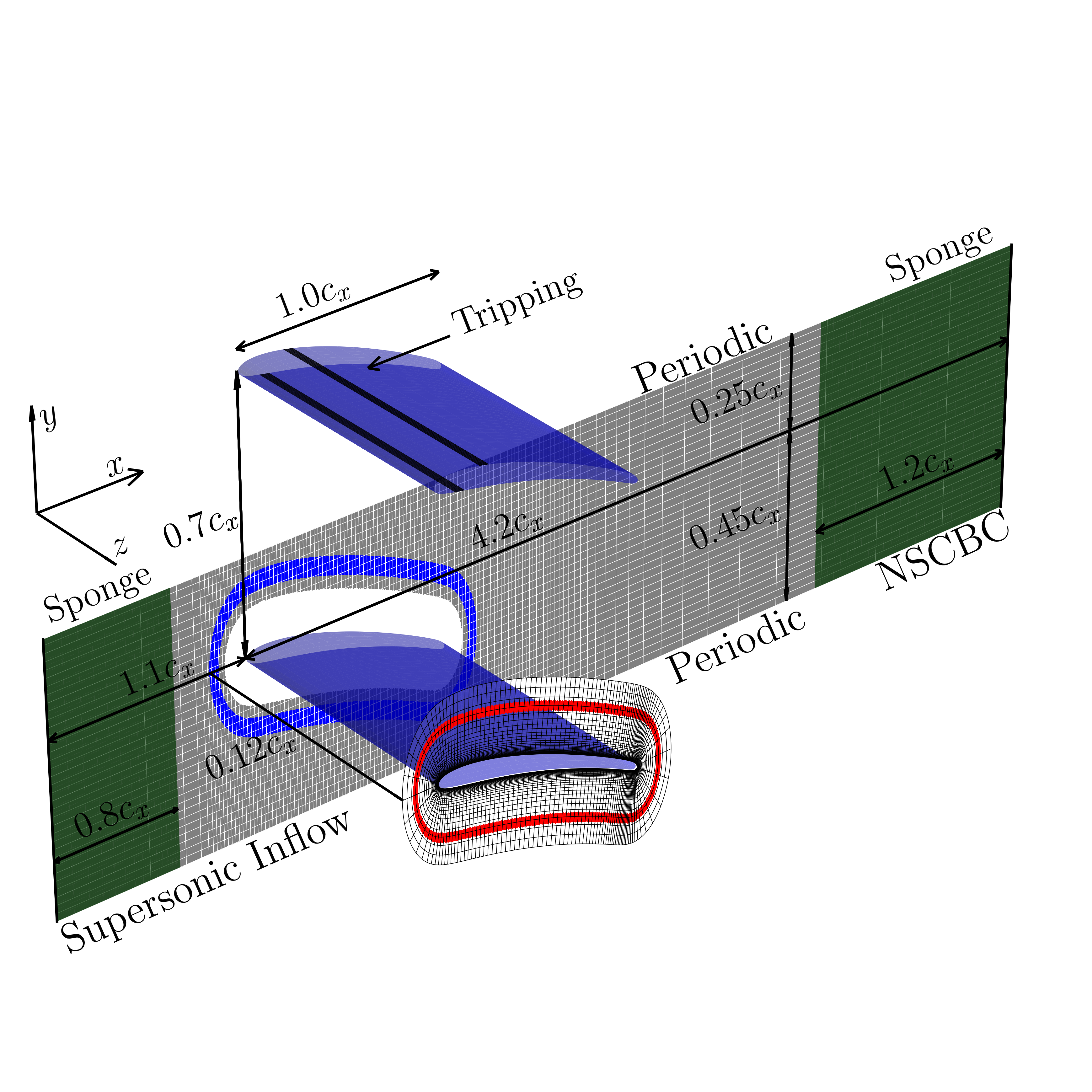}
	\caption{Schematic of the computational domain (skipping every 8 points in all directions) and the boundary conditions. The O-grid is represented by black lines and the Cartesian H-grid by white lines. The red and blue stripes show where the interpolation between the two grid blocks occur, with the information exchange going from O-grid to Cartesian and Cartesian to O-grid, respectively. The wider green stipes show where the damping sponges are applied, and the black stripes on the airfoil surface show where the tripping is applied.}
	\label{fig:schematic}
\end{figure}

A linear cascade of supersonic turbines is solved numerically by wall-resolved LES to explore the effects of SBLIs, streamwise pressure gradients, and wall thermal boundary conditions on turbulent boundary layers. The present blade geometry was optimally designed for a $M_{\infty}=2.0$ inlet condition by \citet{liu2019characterization}. This Mach number was chosen by the previous authors based on studies that evaluated the performance of detonation combustors for different exhaust nozzles \citep{braun2017unsteady}. The same geometry was investigated by \citet{lui2022unsteadiness, lui2024mach} using high-fidelity simulations considering a blade pitch of $0.7c_x$. More recently, experiments were conducted for the same turbine configuration by \citet{namatsaliuk2025experimental}. Details regarding the computational domain can be found in Fig.\ \ref{fig:schematic}. The Reynolds number based on the inlet velocity and airfoil axial chord is $Re_{\infty} = 200,000$ and the inlet Mach number is $M_\infty=2.0$. Adiabatic and isothermal walls are investigated where the latter has a cooled surface with wall temperature $T_w = 0.75T_\infty$. The fluid is a calorically perfect gas with the ratio of specific heats given by $\gamma=1.31$. The Prandtl number is $Pr=0.747$, and the non-dimensional Sutherland constant is $S_\mu = 0.07182$.

The computational domain is composed by two grid blocks, being one O-grid block surrounding the blade, designed to resolve the boundary layer, and a background Cartesian H-grid which facilitates the implementation of the boundary conditions. The O-grid contains $1200\times280\times144$ points in the streamwise ($\xi^1$), wall-normal ($\xi^2$) and spanwise ($\xi^3$) directions, respectively. A hyperbolic grid generation approach is employed to assure that the $\xi^1$ and $\xi^2$ grid lines are locally orthogonal on the O-grid (the $\xi^3$ direction is uniform). The H-grid has $960\times280\times72$ points in the $x$, $y$ and $z$ directions, respectively. Considering both grids, the total number of points is approximately $68$ million. The computational domain and both grid blocks can be seen in Fig.\ \ref{fig:schematic}, where the O and H grids are represented by black and white lines, respectively. Only every 8-th grid point in each direction is shown for visualization purposes. A hole appears on the H-grid along the region where the O-grid is resolved. Since the outer/inner boundary conditions of the O/H grids are not defined, a fourth-order Hermite interpolation scheme is utilized to exchange data between the two grid blocks in the overlapping zones \citep{delfs2001overlapped,bhaskaran2010large}. These overlapping regions are represented by red and blue stripes in Fig.\ \ref{fig:schematic}, representing the region where the information is exchanged between O$\to$H-grid and H$\to$O-grid, respectively.

On the H-grid, a supersonic inflow is used to set the inlet boundary condition, and the Navier-Stokes characteristic boundary condition (NSCBC) is used for the outflow \citep{poinsot1992boundary}. Periodic boundary conditions are imposed on the $y$-direction, on the background grid, to simulate a linear cascade of blades. Periodic boundary conditions are also applied on the $z$-direction for both grid blocks to enforce a spanwise homogeneous flow. For the O-grid, adiabatic and isothermal boundary conditions are employed at the blade surface (shown in blue in Fig.\ \ref{fig:schematic}). In addition to the boundary conditions, damping sponges are applied at the inflow and outflow boundaries of the H-grid to absorb convected numerical disturbances \citep{nagarajan2003robust}. The sponge regions are represented by wide green stripes in Fig.\ \ref{fig:schematic}.

The spatial discretization of the governing equations is performed by a sixth-order accurate compact finite-difference scheme in a staggered grid configuration \citep{nagarajan2003robust}. To obtain the properties in the staggered arrangement, a sixth-order compact interpolation scheme proposed in the previous reference is also used. An implicit second-order time integration scheme is used in the O-grid to overcome the numerical stiffness that arises from the fine boundary layer grid \citep{beam1978implicit}. The flow solution on the H-grid is integrated by a third-order Runge-Kutta scheme \citep{Akselvoll1996Efficient}. To preserve stability of the numerical simulation, a sixth-order compact filter \citep{lele1992compact} is applied to control high-wavenumber numerical instabilities that may arise from mesh curvature and stretching, numerical interpolation, and boundary conditions. The compact filter has two free parameters, $\alpha$ and $\beta$, which control the cutoff wavenumber. It should be mentioned that $\alpha$ can be varied in the range $-0.5 \leq \alpha \leq 0.5$, where the filter is turned off for $\alpha=0.5$. In the present work, $\beta = 0$ and $\alpha$ is linearly varied between 0.5 and 0.45 in the wall-normal direction, starting on the buffer layer until the limit of the O-grid. This means that the filter should only damp high wavenumbers, having less impact along the boundary layer. No explicit subgrid-scale model is applied in the present LES, which follows the approach of \citet{mathew2003explicit}, where high-order compact filters are used to control numerical instabilities without explicitly modeling subgrid-scale effects.

A shock capturing scheme is also employed to introduce numerical dissipation in the vicinity of shock waves without damping the small scales of turbulence. Here, the localized artificial diffusivity scheme LAD-D2-0 from \citet{kawai2010assessment} is used to compute the artificial bulk viscosity and thermal conductivity, which are added to their physical counterparts. 

A realistic supersonic turbine is exposed to inflow perturbations that arise from a detonation combustor. However, a uniform inlet flow is considered in this work. This choice of numerical setup allows a more direct comparison with other canonical flow configurations involving supersonic boundary layers and SBLIs available in the literature. The simulation is initialized with a uniform flow and the transient regime is discarded before collecting statistics. Spanwise-averaged flowfields are saved every $\Delta t = 0.002$, while for 3D flow data we have $\Delta t = 0.006$. Statistics are collected for approximately 36 non-dimensional time units and the averages are evaluated using the whole temporal and spatial domain. In the present case, to promote transition of the boundary layers, an artificial body-force tripping ($f^{i}_{b}$ from Eqs.\ \ref{eq:momentum_equation} and \ref{eq:energy_equation}) is applied near the blade surface on both sides of the airfoil. The unsteady tripping changes every $\Delta t \approx 0.006$ exciting several spanwise wavenumbers in a random fashion \citep{Sansica,waindim2016body}, and the amplitudes of the perturbations are chosen experimentally to guarantee a bypass transition with minimal flow disturbance. The body-force tripping is applied at $0.22< x < 0.27$ on the suction side and $0.10 < x < 0.15$ on the pressure side. These regions are highlighted as black stripes in Fig.\ \ref{fig:schematic}. The wall-normal height of the body-forcing is $0.1\%$ of the axial chord length. 

The spanwise domain is $0.12 c_x$, being around 6 times larger than the incoming boundary layer thickness on the suction side for both cases. This is similar or larger than most high-fidelity computations of SBLIs \citep{loginov2006large,touber2009large,agostini2012zones,adler2018dynamic,lui2024mach}. The near-wall grid spacing in terms of wall units does not exceed $\Delta s^+\approx 60$, $\Delta n^+ \approx 0.6$ and $\Delta z^+ \approx 20$, where $s$, $n$ and $z$ stand for streamwise, wall-normal and spanwise directions, respectively. The largest values of $\Delta^+$ are obtained for the cooled wall case, and such values are obtained away from the tripping and separation regions. Further details about the grids, including spanwise correlations, grid resolution in wall units, and an estimated Kolmogorov scale can be found in \ref{sec:appendixWallUnits}. The present LES tool has been validated for simulations of compressible flows \citep{nagarajan2004,wolf2012convective} including studies of transonic and supersonic turbine cascades \citep{bhaskaran2010large,lui2022unsteadiness,lui2024mach}.

\section{Results}\label{sec:results}

This section presents the characterization of turbulent boundary layers computed for adiabatic and isothermal (cooled) blades. The impact of adverse and favorable pressure gradients, as well as that from SBLIs is  assessed. General features of the investigated supersonic turbine cascade are shown first, followed by the characterization of the developing boundary layers. Then, the effects of pressure gradient, SBLI, and thermal boundary conditions on mean flow properties are investigated. 

All results are presented in non-dimensional form. Instantaneous velocity components in Cartesian coordinates are represented by $U, V, W$ for the $x, y, z$ directions, respectively. The same logic is used to address the instantaneous velocity components in contravariant coordinates, with $u^1 = u_t$, $u^2 = u_n$, and $u^3 = W$ for the streamwise/wall-tangential, wall-normal, and spanwise directions, $\xi^1, \xi^2, \xi^3$, respectively. It is recalled that, for the present homogeneous flow configuration, $\xi^3 = z$. Unless otherwise specified, contravariant velocity components will be employed in the following analysis. The terms ``streamwise" and ``tangential" will be used interchangeably throughout the text. Moreover, flow quantities are evaluated by employing the Reynolds ($f = \overline{f} + f'$) and Favre decompositions ($f = \tilde{f} + f''$), with $\overline{f}$/$\tilde{f}$ and $f'$/$f''$ representing their mean and fluctuation values, respectively.

\subsection{General features of the supersonic turbine cascade}
\label{subsec:generalfeatures}

\begin{figure}[H]
	\centering
	\includegraphics[trim={0.0cm 0cm 0cm 0cm},clip,width=0.99\textwidth]{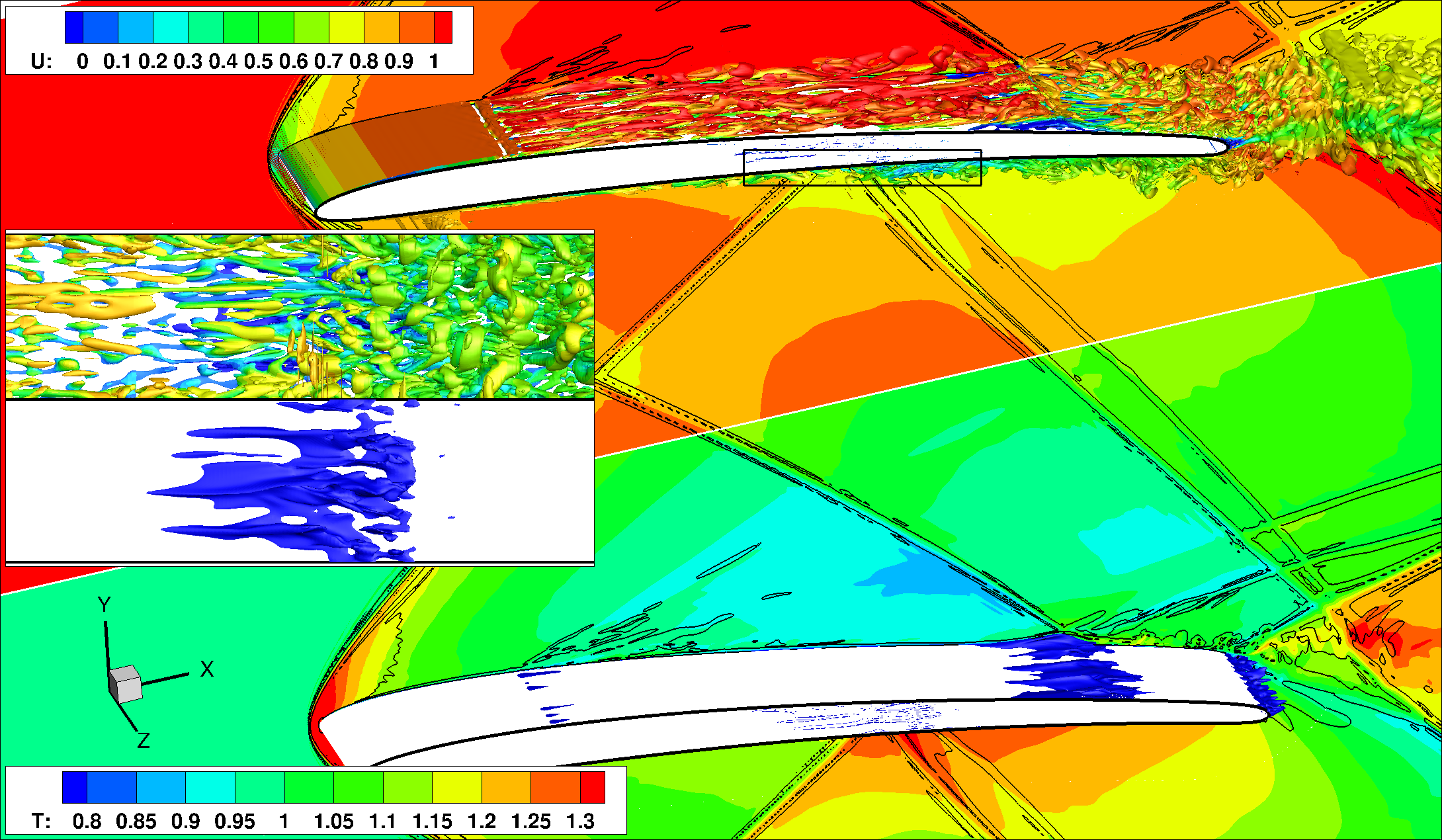}
	\caption{Instantaneous view of the cooled turbine cascade. The top half plot shows a background plane of $U$ velocity plus iso-surfaces of $Q$-criterion ($Q = 10$) colored by $U$. The bottom half plot displays contours of $T$ with an iso-surface of $U=0$ highlighting the flow separation regions. Black lines in the background plane depict the shock waves by visualizing the density gradient magnitude ($|\nabla \rho|$). The inset shows iso-surfaces of $U=0$, highlighting the separation region on the pressure side, and $Q$-criterion colored by $U$.}
	\label{fig:instant_flow}
\end{figure}

A snapshot of the cooled case is shown in Fig.\ \ref{fig:instant_flow} to provide an overall picture of the flow. The figure is separated in two parts: the top half shows instantaneous iso-contours of $U$ velocity in the background plane, with iso-surfaces of $Q$-criterion for $Q=10$ colored by $U$ to highlight the near-wall turbulent structures. The bottom half shows instantaneous iso-contours of temperature $T$ in the same plane, with an iso-surface of $U = 0$ along the airfoil surface to highlight the regions of separated flow. Both plots are split by a white line that denotes the periodic boundary condition in the $y$-direction. The inset is a bottom-up view of the region represented by the black line rectangle, and shows an instantaneous view of the separation bubble on the pressure side, together with turbulent structures depicted by iso-surfaces of $Q$-criterion. These structures are colored by the same contour levels of $U$ as shown for the suction side. The black lines in the background plane represent iso-contours of density gradient magnitude ($|\nabla \rho|$), enabling the visualization of shock waves and shocklets. Detached shocks form at the leading edges of the blades. They impinge on the suction side of the neighboring blades as an oblique shock, whereas a Mach reflection forms on the pressure side.

The SBLIs induce separation bubbles, which are visualized using iso-surfaces of $U = 0$ on both the suction and pressure sides of the airfoil. These separation bubbles clearly exhibit three-dimensional structures, despite the spanwise-periodic nature of the flow. On the suction side, the undulating shape of the bubble arises from the presence of high- and low-speed streaks upstream of the SBLI, which penetrate the bubble and lead to smaller and larger separation regions, as discussed by \citet{lui2022unsteadiness}.

The iso-surfaces of $Q$-criterion upstream of the SBLIs reveal that the near-wall structures are predominantly streamwise-elongated. These structures arise due to two main factors. The first is the flow’s response to disturbances introduced by the boundary layer tripping. \citet{lui2022unsteadiness} showed that such structures emerge as a natural response to spanwise-random flow disturbances, and the employed tripping device does not directly excite the wavenumbers of the elongated structures. In the present case, tripping induces a bypass transition in the boundary layer, characterized by the algebraic growth of disturbances in the form of streamwise-elongated streaks. Similar structures were reported by \citet{bugeat20193d}, who performed a bi-global resolvent analysis of a supersonic boundary layer over a flat plate and found that the optimal response modes resemble streaky structures of the same nature. The second contributing factor is the relatively low Reynolds number and limited streamwise development, which inhibit the boundary layer from fully transitioning into a turbulent state. The combination of these two effects results in the observed streamwise-elongated structures upstream of the SBLIs.

The turbulent structures become finer downstream of the SBLIs, and exhibit stronger three-dimensionality. \citet{volpiani2018effects} observed a similar behavior downstream the SBLI, although their boundary layer was already fully turbulent before the interaction. It is worth noting that the same flow features are observed for the adiabatic blade configuration, and are not shown here for brevity.

\begin{figure}[H]
	\centering
	\includegraphics[trim={0.0cm 0cm 0cm 0cm},clip,width=0.99\textwidth]{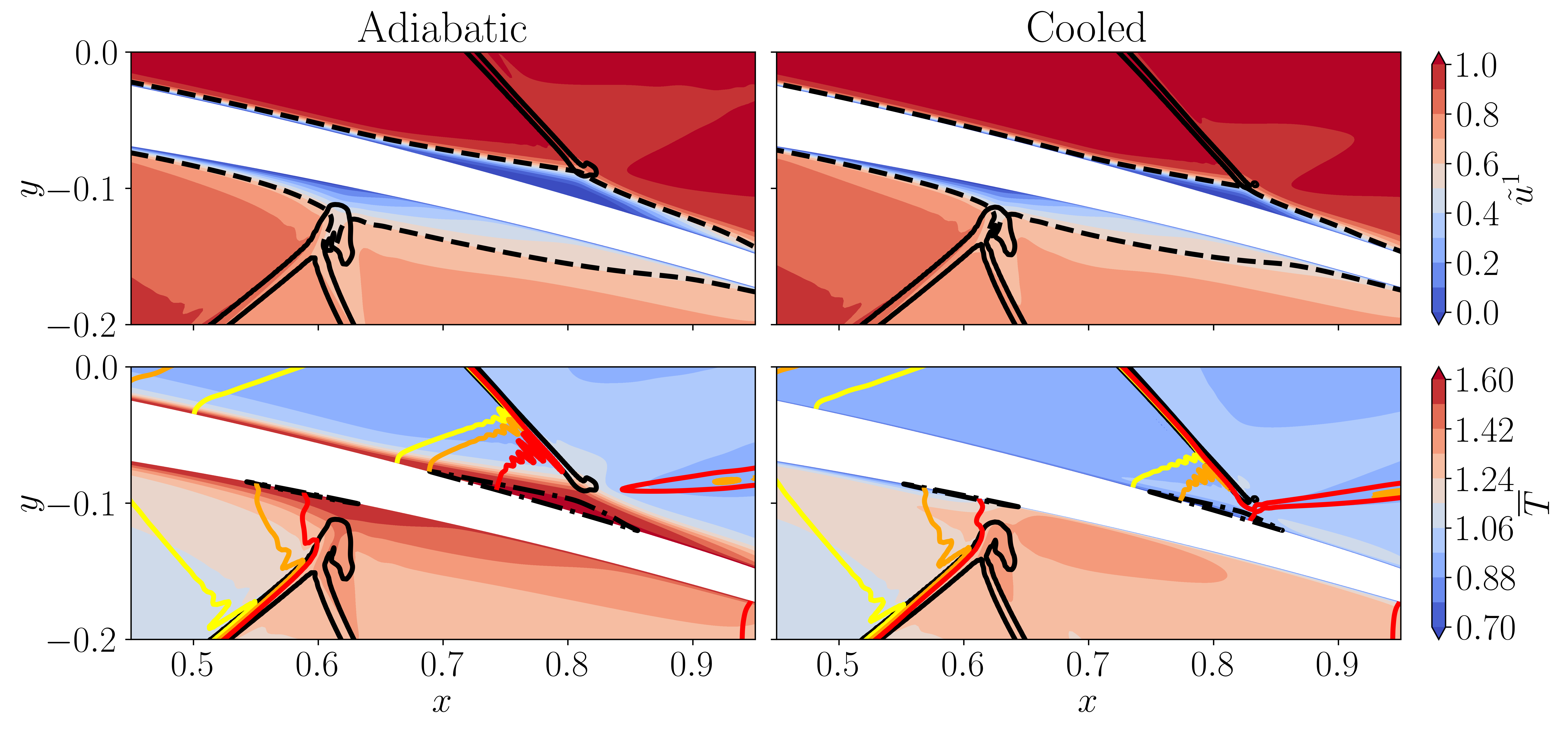}
	\caption{Mean contours of the adiabatic (left) and cooled (right) cases. The first and second rows show iso-surfaces of mean tangential velocity ($\tilde{u}^1$) and temperature ($\overline{T}$), respectively. Black solid iso-lines show the magnitude of the mean density gradient ($|\nabla\overline{\rho}|$), highlighting the shock waves. Dashed lines represent sonic lines (top), and dashed dotted lines delimit the separation regions (bottom). Iso-lines of mean pressure levels are shown by solid color lines (yellow, orange, red) highlighting the compression regions prior to the shock impingement. For the suction side, the iso-lines of mean pressure levels are (yellow, orange, red) = $(0.55,0.58,0.61)$, and for the pressure side (yellow, orange, red) = $(1.5,1.8,2.1)$.}
	\label{fig:mean_overview}
\end{figure}

To assess the impact of the blade thermal boundary conditions on the separation regions, iso-surfaces of mean tangential velocity ($\tilde{u}^1$) and temperature ($\overline{T}$) are shown in the first and second rows of Fig.\ \ref{fig:mean_overview}, respectively. The left column presents solutions of the adiabatic case, whereas the right column shows the cooled one. The sonic line is represented by a black dashed line on the top row, while the separation region can be visualized as a dashed dotted line on the bottom row. Iso-lines of mean pressure can be seen as yellow, orange and red solid lines, representing lower to higher pressure levels. Also, black solid iso-lines of  ($|\nabla\overline{\rho}|$) enable the visualization of shock waves. An analysis of the shock waves in Fig.\ \ref{fig:mean_overview} reveals that there is no permanent reattachment shock on the suction side in either case. In fact, the data indicate that this reattachment shock is intermittent, and its presence can be observed in the instantaneous snapshot shown in Fig.\ \ref{fig:instant_flow}.

When comparing the recirculation regions of the adiabatic and isothermal cases, one can observe that the latter has a smaller separation bubble on the suction side. Furthermore, the sonic line is closer to the wall when cooling is applied, especially closer to the separation region. This phenomenon is caused by the difference in the wall temperature in the two cases. Since the near-wall temperature of the cooled case is lower than that of the adiabatic blade, it leads to a lower local speed of sound, which results in supersonic flow closer to the wall. This in turn leads to a deeper penetration of the shock wave into the boundary layer, visible especially on the suction side. 

The temperature contours exhibit distinct behavior between the two cases. The adiabatic blade shows higher temperatures near the wall, within the boundary layers, compared to the isothermal case. This is due to the absence of heat transfer to the wall. Hence, the friction generated by the shear stresses is converted into heat and advected through the boundary layer. For the cooled surface, lower temperatures are observed near the wall. Downstream of the separation region on the pressure side, the cooled case exhibits an abrupt temperature rise caused by the normal shock.

\begin{figure}[H]
	\centering
	\includegraphics[trim={0.0cm 0cm 0cm 0cm},clip,width=0.99\textwidth]{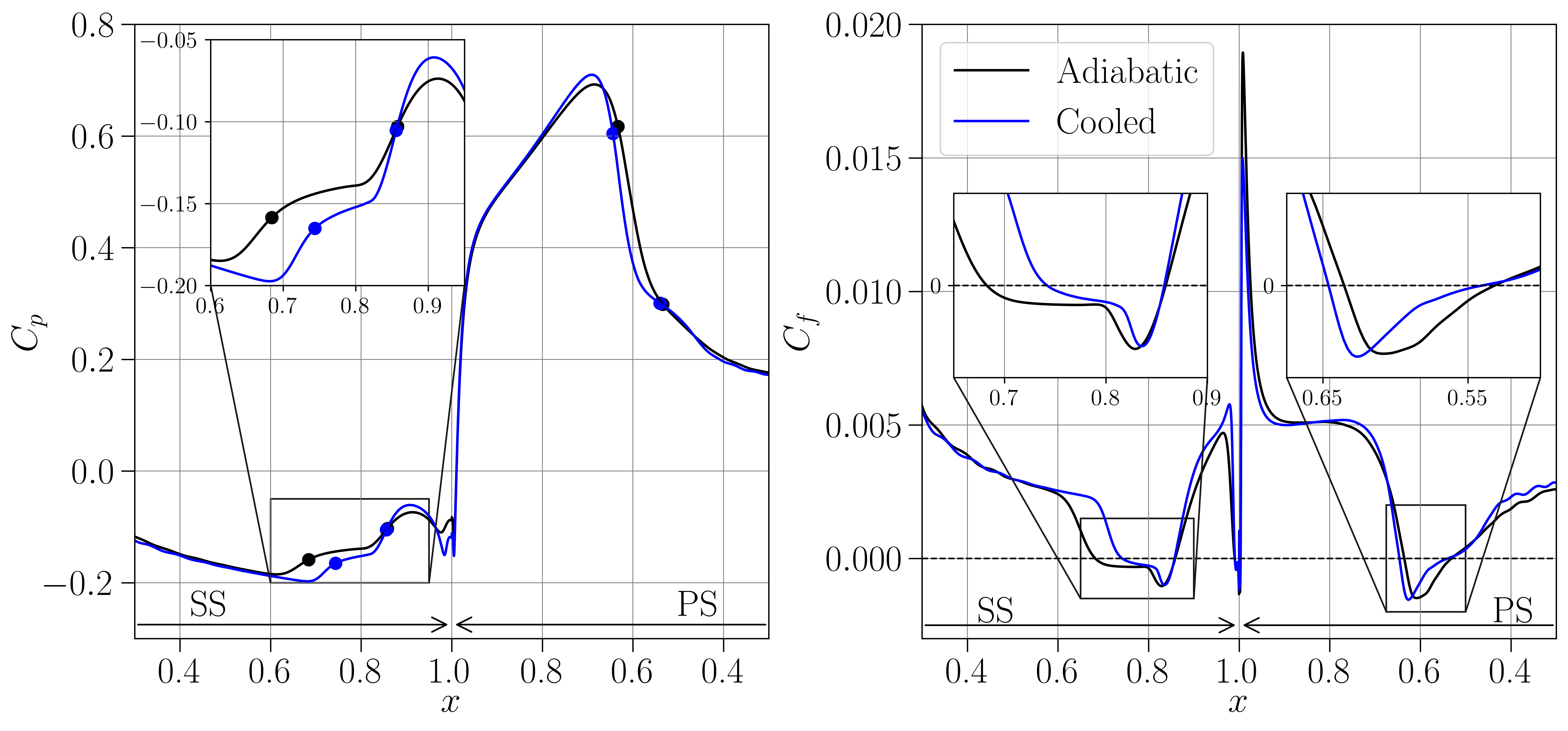}
	\caption{Profiles of mean pressure coefficient (left) and mean skin-friction coefficient (right) for the adiabatic (black) and cooled (blue) blades. The black and blue dots represent the separation and reattachment positions for the adiabatic and cooled cases, respectively.}
	\label{fig:cfcp}
\end{figure}

The mean pressure coefficient $\left(C_p = \frac{\overline{p}_w-p_\infty}{0.5\rho_\infty U_\infty^2}\right)$ and the mean skin-friction coefficient $\left(C_f = \frac{\overline{\mu}(\overline{u}^1_w)_{,2}}{0.5\rho_\infty U_\infty^2}\right)$ are shown along the airfoil chord in Fig.\ \ref{fig:cfcp} to further analyze the separation regions. Here, the subscript $w$ stands for parameters evaluated at the wall. The plots read as follows: the suction side (SS) goes from left to center ($x = 1.0$), whereas the pressure side (PS) goes from right to center. The horizontal dashed line in the $C_f$ plot delimits the separation region, and the black and blue dots in the $C_p$ plot represent the separation and reattachment locations for the adiabatic and cooled cases, respectively.

The $C_p$ distribution shows that the boundary layer on the suction side develops under a favorable pressure gradient, whereas on the pressure side it develops under an adverse pressure gradient. On both sides of the airfoil, two pressure jumps are noticed for both cases. The first is related to compression waves upstream of the separation position, which is more subtle on the suction side than on the pressure side due to a weaker compression, as also shown in Fig.\ \ref{fig:mean_overview}. The second jump is related to the shock impingement. Due to the normal shock, the pressure side displays a stronger jump compared to the suction side. On the other hand, the suction side has a pressure plateau between the rises. The pressure gradients on the second jumps are steeper for the cooled case, regardless of the airfoil side, compared to the adiabatic configuration. This is due to the higher penetration of the shock waves caused by wall cooling (Fig.\ \ref{fig:mean_overview}).

\begin{figure}[H]
	\centering
	\includegraphics[trim={0.0cm 0cm 0cm 0cm},clip,width=0.99\textwidth]{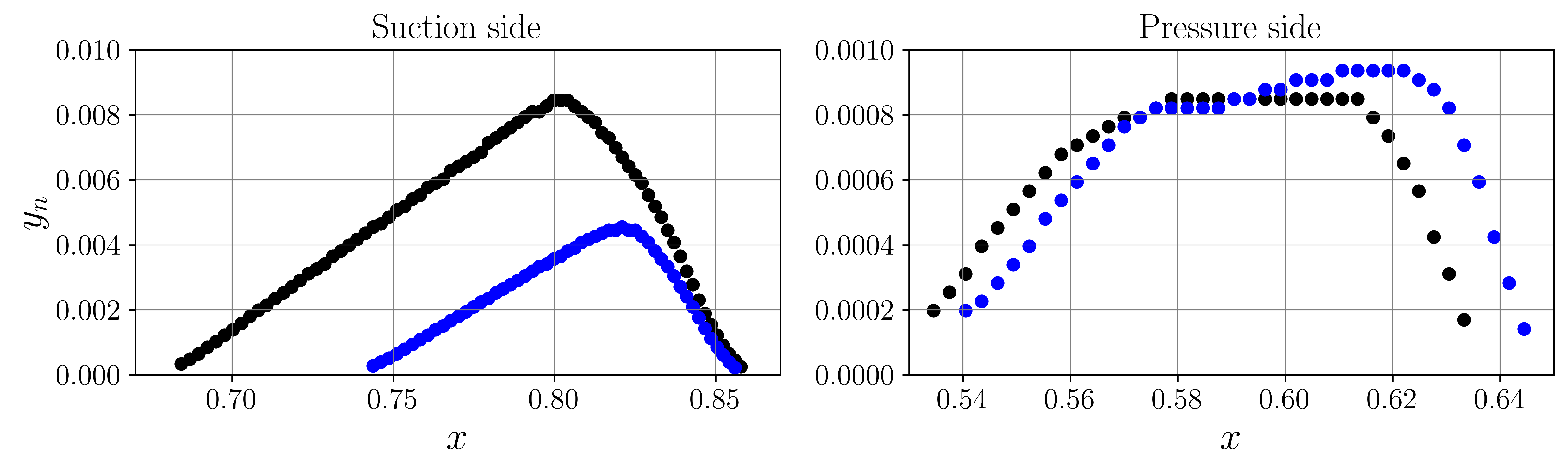}
	\caption{Separation bubble shapes for the isothermal (blue) and adiabatic (black) cases. The left plot shows the suction side bubbles whereas the right one shows the pressure side.}
	\label{fig:bubble_size}
\end{figure}

The $C_f$ plot shows that the flow separates earlier for the adiabatic case on the suction side compared to the isothermal configuration. However, the reattachment location is almost the same for both wall boundary conditions. This leads to a larger separation bubble for the adiabatic case. \citet{volpiani2018effects} observes a similar behavior for DNS of isothermal turbulent boundary layers at $M_{\infty}=2.28$. In this reference, the authors show a smaller deviation of the reattachment point for a cooled wall, whereas a larger deviation is shown for the separation point. On the pressure side of the present turbine, the isothermal case also shows later separation, but it also reattaches downstream. An inspection of the $C_p$ distribution in Fig. \ref{fig:cfcp} shows that the impinging oblique shocks on the suction side are located in the same position for the adiabatic and isothermal walls. On the other hand, the normal shock from the Mach reflection on the pressure side is slightly displaced downstream for the isothermal surface. Based on the present observations, the reattachment points are impacted mainly by the shock positions, while the separation points are affected by the thermal conditions on the suction side and the shock location on the pressure side. A similar observation concerning the reattachment location was also reported by \citet{bernardini2016heat}. Moreover, for strong shock interactions, \cite{volpiani2020effects} observe that, as the flow separates, the $C_f$ distribution shows a small plateau followed by a further drop before reattaching. This behavior is clearly observed on the suction side, but not on the pressure side, despite the stronger shock interaction. This phenomenon could be an effect of the more flattened separation bubbles on the pressure side compared to those on the suction side, visible in Figs.\ \ref{fig:mean_overview} and \ref{fig:bubble_size}.

Figure \ref{fig:bubble_size} shows the mean separation bubble shapes for the suction and pressure sides. The bubble heights are computed in terms of the wall-normal distance $y_n$. One can observe that the pressure side bubbles are one order of magnitude smaller in height than those on the suction side. Furthermore, both separation bubbles have similar lengths and heights on the pressure side whereas, on the suction side, the bubble is considerably smaller and thinner for the isothermal blade. This indicates that the size of the pressure side bubble is not determined solely by the surface thermal boundary condition. Other parameters, such as the pressure gradient and shock interaction strength, also play a significant role in flow separation, as shown by \citet{lui2024mach}. The role of these parameters in the present supersonic boundary layers is investigated in the following sections.

\subsection{Boundary layer characterization}
\label{subsec:boundarylayercharacterization}

A characterization of the boundary layers is provided in this section, starting with the evaluation of the boundary layer thickness $\delta$, followed by integral parameters such as the displacement and momentum thicknesses. The boundary layer thickness is evaluated using the intermittency factor of the instantaneous vorticity magnitude as shown by \citet{jimenez2010turbulent} and \citet{vinuesa2016determining}. In this case, the intermittency of the turbulent/nonturbulent interface is tracked for a vorticity threshold. The robustness of this approach was demonstrated for different Reynolds numbers by the previous authors, who found that the boundary layer thickness would be set for an intermittency factor of $0.3$. The supersonic turbine cascade boundary layer is subjected to fluctuating streamwise pressure gradients besides different SBLIs, and the selected threshold for which convergence of $\delta$ is achieved is $|\omega| = \sqrt{\omega_i\omega_i}= 12.5$. This particular method of boundary layer thickness calculation is chosen compared to the threshold of $z$-vorticity magnitude employed by \cite{pirozzoli2010direct} and \cite{lui2024mach} due to its better resolution downstream of the SBLIs.
\begin{figure}[H]
	\centering
	\includegraphics[trim={0.0cm 0cm 0cm 0cm},clip,width=0.99\textwidth]{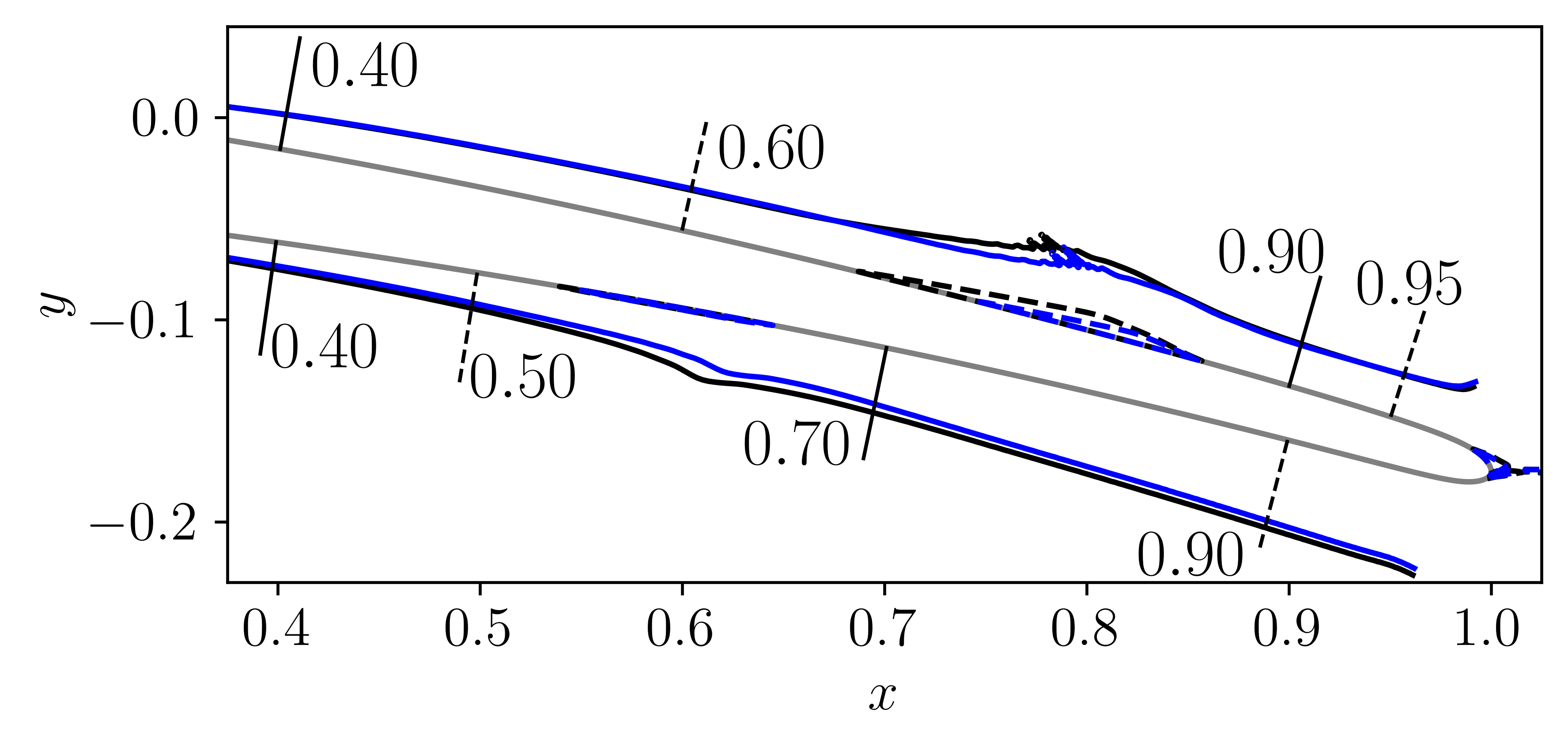}
	\caption{Boundary layer thickness (solid lines) and separation bubbles (dashed lines) of adiabatic (black) and cooled (blue) blades. The multiple wall-normal solid and dashed lines represent the streamwise locations where mean flow quantities are extracted in the following sections.}
	\label{fig:bl_thickness}
\end{figure}

The streamwise variations of $\delta$ on the suction and pressure sides are shown by the blue and black solid lines in Fig.\ \ref{fig:bl_thickness}, for the cooled and adiabatic walls, respectively. One can see that the values of $\delta$ are similar along the airfoil for both thermal boundary conditions, except above the recirculation regions, indicated by dashed lines. Downstream of the SBLIs, the boundary layers thicken due to free shear layers induced by the separation bubbles, and this effect is more evident on the pressure side. Different positions along the airfoil chord in Fig.\ \ref{fig:bl_thickness} are highlighted in which data are extracted in the wall-normal direction to assess, in the following sections, the impact of pressure gradient, SBLI, and thermal boundary conditions on the spatial evolution of mean flow quantities.

\begin{figure}[H]
	\centering
	\includegraphics[trim={0.0cm 0cm 0cm 0cm},clip,width=0.99\textwidth]{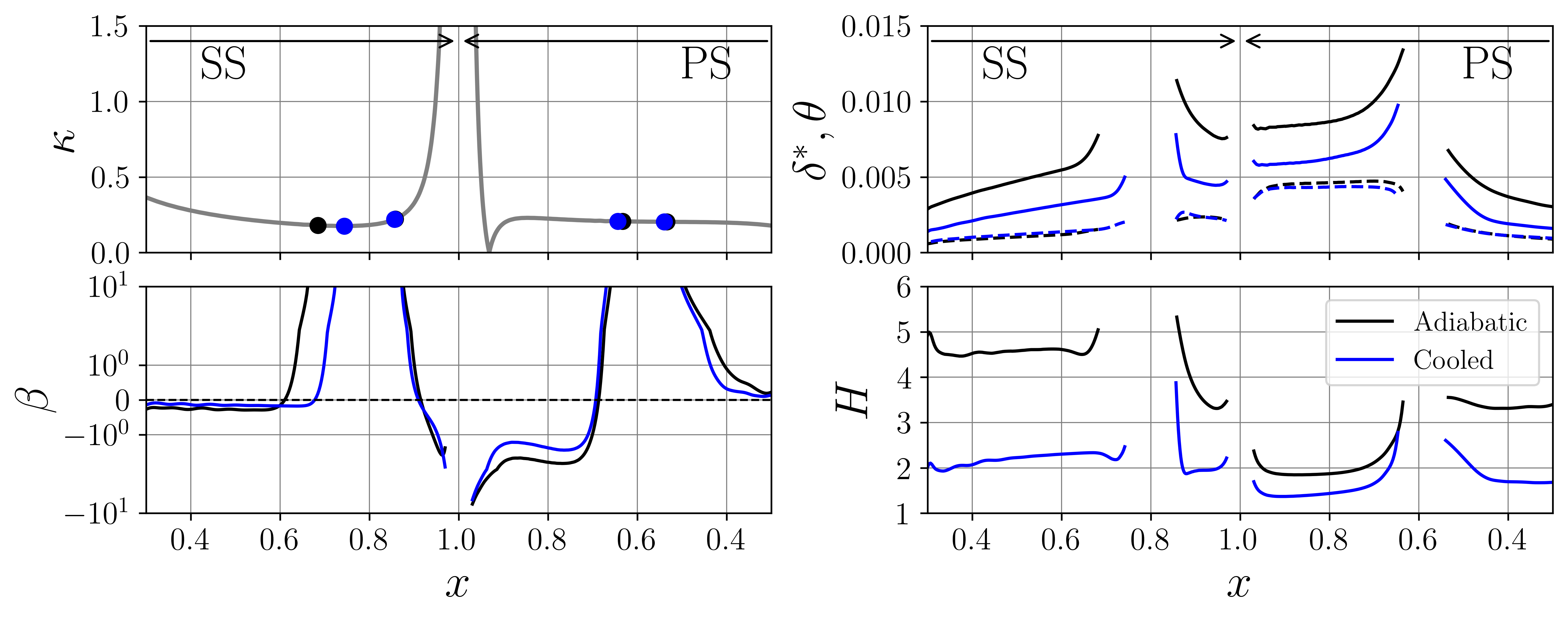}
	\caption{The first row shows the local curvature (left) and the displacement (solid lines) and momentum (dashed lines) thickness distributions over the blades (right). The second row shows the Clauser parameter (left) and the shape factor (right). Black/blue lines are used for the adiabatic/cooled cases. In the curvature plot, the black/blue dots stand for the separation and reattachment positions of the adiabatic/cooled cases.}
	\label{fig:integral_parameters}
\end{figure}

Figure \ref{fig:integral_parameters} presents a detailed characterization of the boundary layers as it depicts the local surface curvature ($\kappa = 1/R$), besides chordwise distributions of integral parameters such as displacement thickness ($\delta^* = \int_{0}^{\delta} [1-(\overline{\rho}/\overline{\rho_e})(\overline{u}^1/\overline{u}^1_e)]dy_n$), and momentum thickness ($\theta = \int_0^{\delta}(\overline{\rho}/\overline{\rho_e})(\overline{u}^1/\overline{u}^1_e)[1-(\overline{u}^1/\overline{u}^1_e)]dy_n$). The former integral parameter is a measure of the streamline displacement due to the presence of a boundary layer, and the latter is related to the friction drag. Furthermore, Fig.\ \ref{fig:integral_parameters} displays the Clauser parameter ($\beta = [\delta^*/\overline{\mu}(\overline{u}_w^1)_{,2}](\overline{p}_w)_{,1}$) (\cite{clauser1954turbulent}), which characterizes the local pressure gradient, including flow history effects, and the shape factor ($H = \delta^*/\theta$), which is related to the likelihood of boundary layer separation. In the $\delta^*$ and $\theta$ equations, the subscript $e$ stands for quantities evaluated at the edge of the boundary layer.

The present airfoil profile has convex and concave surfaces on the suction and pressure sides, respectively, and the local surface curvature, $\kappa$, is shown in Fig.\ \ref{fig:integral_parameters}. The same dot markers indicated in Fig.\ \ref{fig:cfcp} are displayed in Fig.\ \ref{fig:integral_parameters} to highlight the separation and reattachment positions. The boundary layers and bubbles develop in regions of small curvature, away from the blade leading and trailing edges. Therefore, this parameter is not expected to significantly impact the development of the boundary layer for the present configuration. However, surface curvature leads to other effects that impact the boundary layers, such as variations in the streamwise pressure gradient.

The local pressure gradient is assessed by the Clauser parameter $\beta$ in Fig.\ \ref{fig:integral_parameters}. Upstream of the separation regions, $\beta$ indicates negative and positive values on the suction and pressure sides, respectively. Hence, the boundary layers develop under favorable and adverse pressure gradients, although the magnitudes of $\beta$ are small along these regions. On the separation regions, a sharp increase in the Clauser parameter is observed due to the compression waves on both sides of the airfoil (see Fig.\ \ref{fig:cfcp}). Downstream of the SBLIs, $\beta$ changes drastically from a strong adverse pressure gradient to a moderate favorable pressure gradient condition. On the suction side, this phenomenon occurs near the trailing edge so that the boundary layer does not have sufficient time to develop under the FPG. On the other hand, the pressure side boundary layer has sufficient time to develop under a favorable pressure gradient due to the earlier reattachment. This can be seen in the negative plateaus of $\beta$ in the figure.

Considering the different pressure gradients along the airfoil, the present boundary layers can be classified in four cases: 1) developing on the suction side, upstream of the SBLI, under a favorable pressure gradient (S-U-F), 2) developing on the suction side, downstream of the SBLI, mostly under a favorable pressure gradient (S-D-F), 3) developing on the pressure side, upstream of the SBLI, under an adverse pressure gradient (P-U-A), and 4) developing on the pressure side, downstream of the SBLI, under a favorable pressure gradient (P-D-F). The abbreviations defined above can be interpreted as follows: the first letter refers to the suction/pressure side (S/P), the middle letter informs the upstream/downstream position with respect to the SBLI (U/D), and the last letter corresponds to the pressure gradient imposed on the boundary layer, either favorable or adverse (F/A).

The integral parameters $\delta^*$ and $\theta$ are also displayed in Fig.\ \ref{fig:integral_parameters}. Regardless of the airfoil side, $\delta^*$ grows in the flow direction with a sharp rise near the separation point, followed by a rapid reduction after the reattachment point. This sharp rise and decrease is expected due to the SBLI. After the effect of the SBLI, in the P-D-F region, the value of $\delta^*$ decreases until the trailing edge. This occurs because the streamwise velocity profile becomes fuller closer to the wall as it develops, which is not expected since the boundary layer is subjected to a favorable pressure gradient \citep{spina1994physics}. This behavior indicates that the pressure gradient alone is not responsible for changes in the boundary layer, and this discussion is developed further in the following sections. One can observe a mild increase of the $\theta$ values on both sides of the airfoil, upstream of the boundary layer separation. Downstream of the reattachment point of the boundary layer, the momentum thickness is insensitive to the chord position, being almost constant independently of the thermal boundary conditions for both sides of the airfoil.

It is clear from Fig.\ \ref{fig:integral_parameters} that the thermal boundary conditions have an important effect on the displacement thickness. For the adiabatic case, the higher near-wall temperature (see Fig.\ \ref{fig:mean_overview}) reduces the fluid density near the blade, leading to mass flux towards the outer region \citep{spina1994physics}. This effect should thicken the boundary layer compared to the cooled case, but this is not what is observed in Fig.\ \ref{fig:bl_thickness}. This occurs because the definition of the boundary layer thickness in the current work is purely kinematic, and it does not take into consideration the thermodynamic properties. However, even though the boundary layer thickness is almost identical, a considerable difference is observed for the displacement thickness between the adiabatic and isothermal cases, with the cooled blade depicting smaller values along the chord. This is explained by surface cooling, which not only increases the near-wall density but also reduces the viscosity which, in turn, increases the local Reynolds number. This leads to a fuller velocity profile with a lower displacement thickness throughout the domain, when compared to the adiabatic case.

The shape factor $H$ is also presented in Fig.\ \ref{fig:integral_parameters} to quantify the fullness of the velocity profile, where a higher $H$ indicates a boundary layer more prone to separation. Upstream of the SBLI, the suction side shape factors have values of $H \approx 2$ and $4.5$ for the isothermal and adiabatic surfaces, respectively. Lower values of $H$ are observed on the pressure side, where $H \approx 1.8$ for the isothermal case and $3.3$ for the adiabatic case. Comparable values of shape factors are observed in the literature. For an adiabatic flat plate, \citet{bernardini2016heat} computes $H=3.64$ for a turbulent boundary layer with freestream Mach number $M_{\infty}=2.28$. \cite{cogo2023assessment} presents shape factors of $H=2.29$ to $2.98$ for cooled and adiabatic flat plates, respectively, for turbulent boundary layers at $M_{\infty}=2.0$.

Regardless of the side of the airfoil, adiabatic surfaces always show higher values of $H$ compared to isothermal surfaces. This is consistent with the flowfield observed on the suction side, for which the separation of the boundary layer occurs further upstream for the adiabatic case. For the pressure side, the adiabatic case is also more prone to separate compared to the isothermal wall, but both separation points occur nearly at the same location. Moreover, the pressure side bubbles have similar sizes, differently from the suction side bubble. This corroborates that another mechanism, other than the thermal boundary condition, is responsible for dictating the separation position and bubble size on the pressure side. After reattachment of the boundary layer, on both sides of the airfoil, $H$ is smaller than upstream of the separation point. This occurs because of the energization of the boundary layers due to the SBLIs, which lead to fuller velocity profiles compared to upstream of the separation point, making the boundary layers less prone to separate. A similar behavior in terms of the shape factor reduction is observed by \citet{quadros2018numerical} downstream of the SBLI in a transitional supersonic boundary layer at $M_{\infty}=1.7$.

\subsection{Evolution of mean flow properties along boundary layers}
\label{subsec:effectsinmeanparameters}

This section presents an analysis of the variations in mean flow properties along the boundary layers. Results are shown along the wall-normal direction at the positions depicted by solid and dashed black lines in Fig.\ \ref{fig:bl_thickness}. In the following figures, the solid/dashed lines represent the upstream/downstream positions along the blade, as indicated in the individual subplots. Black and blue lines represent solutions computed for adiabatic and isothermal boundary conditions, respectively. These relations are maintained for the different flow regions analyzed as S-U-F, P-U-A, S-D-F, and P-D-F.  

\subsubsection{Analysis of thermodynamic properties} \label{subsubsec:thermodynamic_properties}

\begin{figure}[H]
	\centering
	\subfigure[Mean density]{\includegraphics[trim={0.0cm 0cm 0cm 0cm},clip,width=0.99\textwidth]{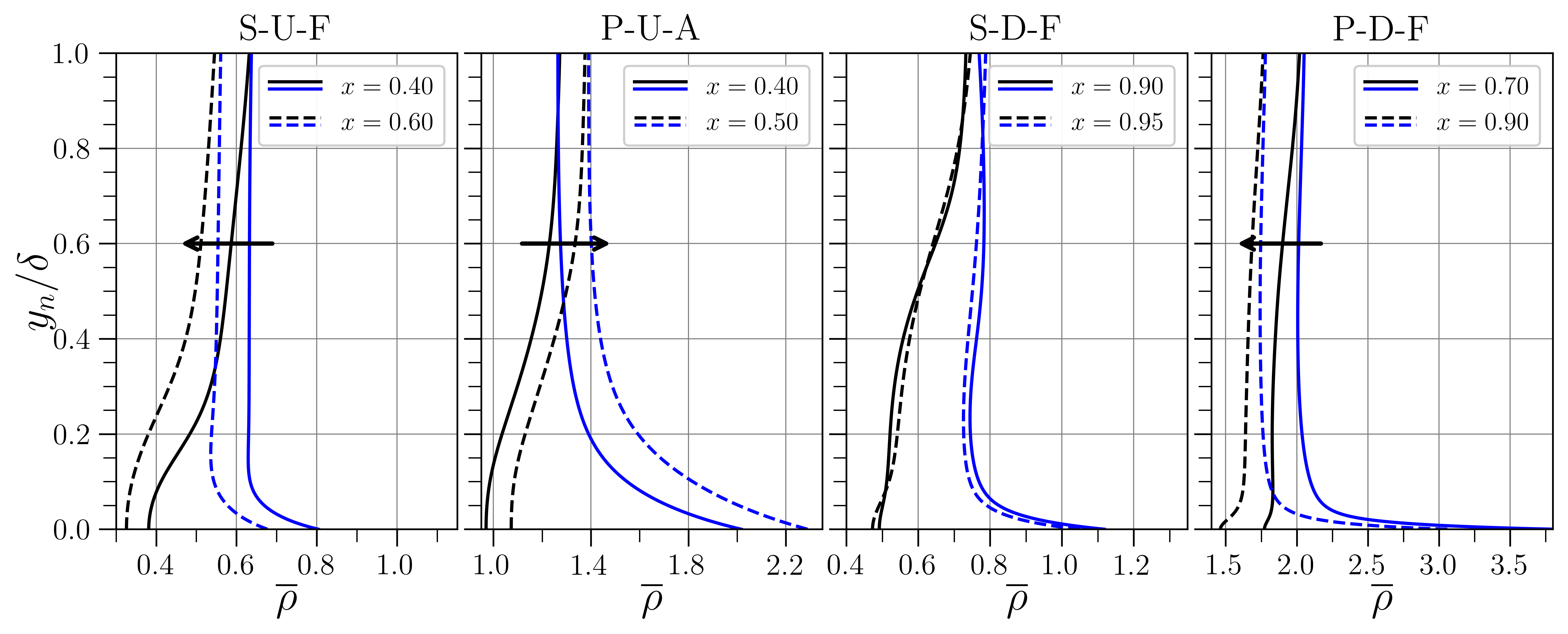}\label{fig:density}}
    \subfigure[Mean temperature]{\includegraphics[trim={0.0cm 0cm 0cm 0cm},clip,width=0.99\textwidth]{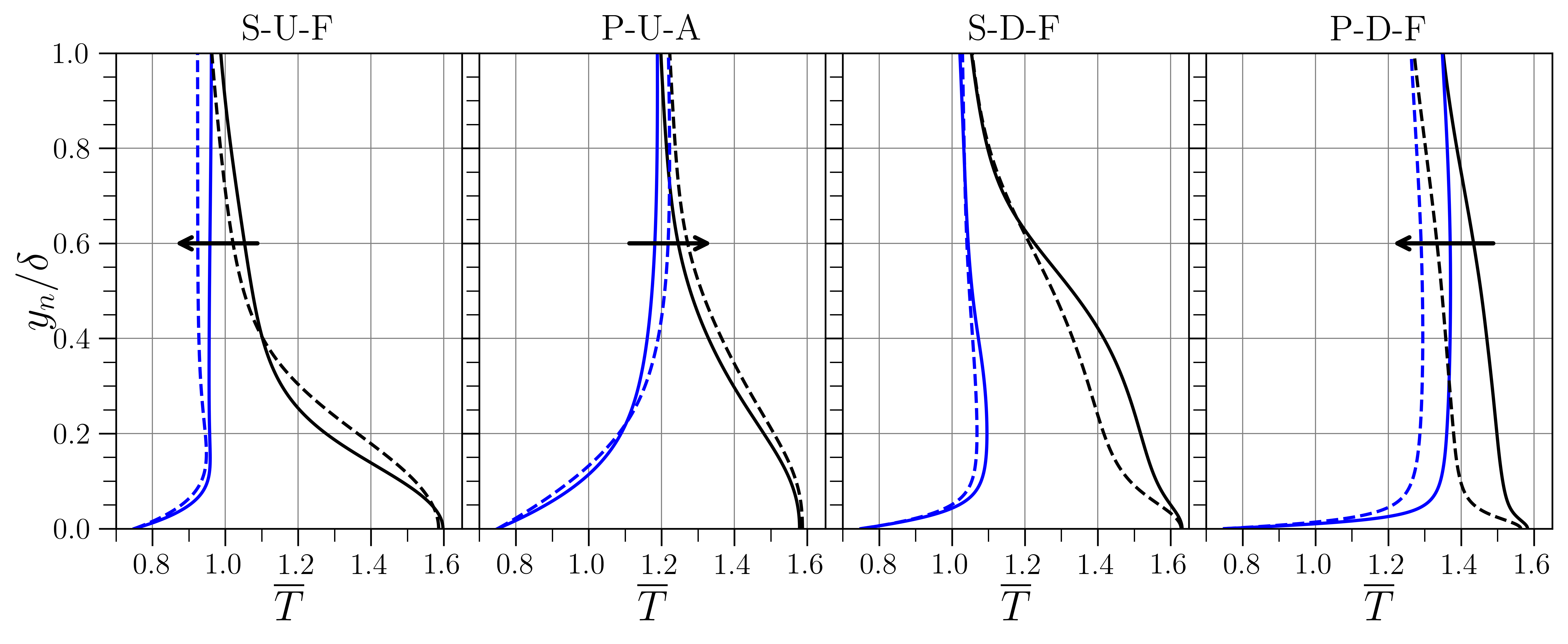}\label{fig:temperature}}
	\caption{Density and temperature profiles for adiabatic (black lines) and isothermal (blue lines) cases for the locations specified in Fig.\ \ref{fig:bl_thickness}. From left to right, the four locations are S-U-F, P-U-A, S-D-F, P-D-F. The solid/dashed lines represent the upstream/downstream positions, respectively.}
    \label{fig:thermodynamic_profiles}
\end{figure}

Figure \ref{fig:thermodynamic_profiles} shows variations in mean density and temperature profiles. These data allow a comparison of thermodynamic quantities for the suction and pressure sides, upstream and downstream of the SBLIs. A comparison of the pressure gradient effects on the solutions reveals that a flow expansion is induced by the favorable pressure gradient (S-U-F and P-D-F), resulting in a progressive decrease in the mean density and temperature profiles as the boundary layers develop in the streamwise direction. In contrast, the presence of an adverse pressure gradient (P-U-A) leads to flow compression, which increases which increases both density and temperature. Notably, the pressure gradient exerts a stronger influence on the density profile than on the temperature profile, particularly upstream of the SBLIs. These trends are highlighted by black arrows in the corresponding figures.

The impact of pressure gradient on the density and temperature profiles is observed throughout the entire boundary layer, with the exception of the S-D-F case. In this region, the flow changes from a high adverse pressure gradient to a high favorable pressure gradient over a short distance (see Fig.\ \ref{fig:integral_parameters}). As a result, the density and temperature profiles remain similar since the flow does not have enough space to adapt to changes in the pressure gradient (see Fig.\ \ref{fig:cfcp}). 

The influence of thermal boundary conditions on the mean density and temperature profiles is also noteworthy. The adiabatic case shows lower densities and higher temperatures near the wall compared to the isothermal case due to heat transfer to the fluid instead of the blade. Another important observation is that, changing the boundary condition from adiabatic to isothermal leads to an opposite behavior in terms of density and temperature gradients in the wall-normal direction. Away from the wall, the density increases for the adiabatic boundary condition, while it decreases for the isothermal configuration. The opposite behavior is observed for the temperature profiles.

\subsubsection{Analysis of streamwise momentum profiles} 
\label{subsubsec:momentum_profiles}

\begin{figure}[H]
	\centering
	\includegraphics[trim={0.0cm 0cm 0cm 0cm},clip,width=0.99\textwidth]{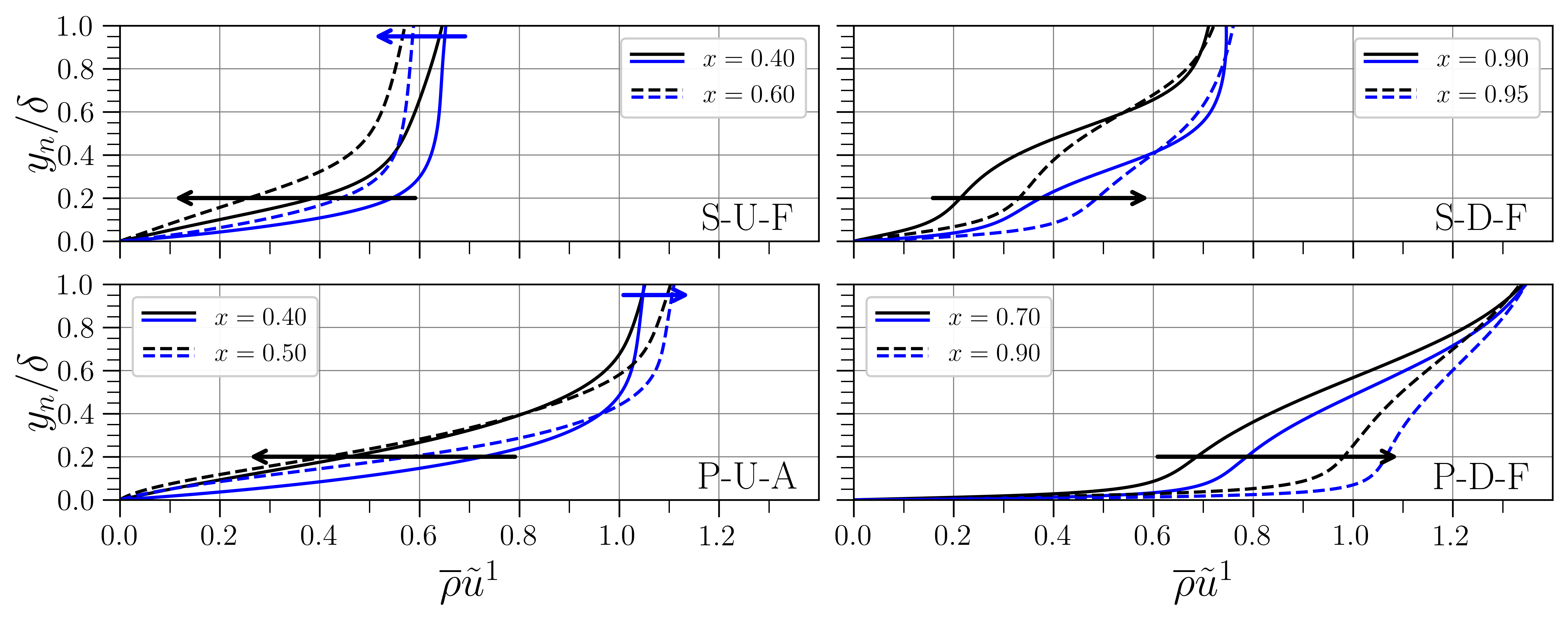}
	\caption{Tangential momentum profiles for the suction side (top) and pressure side (bottom), upstream of SBLI (left) and downstream of SBLI (right), with black and blue lines representing the adiabatic and cooled cases, respectively.}
	\label{fig:momentumx}
\end{figure}

An assessment of streamwise momentum ($\rho u^1$) profiles is presented for different flow positions along the blade in Fig.\ \ref{fig:momentumx}. Data are presented in terms of Favre-averaged quantities $\overline{\rho} \tilde{u}^1$. The streamwise momentum calculated in the isothermal case is always fuller than the one of the adiabatic case, at the same chord location. This behavior indicates a more energized boundary layer and lower chances of separation. This is corroborated by the shape factor analysis of Fig.\ \ref{fig:integral_parameters}.

An analysis of the suction side boundary layer, upstream of the SBLI (S-U-F), shows that the streamwise momentum is reduced at a downstream position for both the isothermal and adiabatic cases. This occurs despite the favorable pressure gradient. Indeed, the streamwise velocity profiles (see Fig.\ \ref{fig:ut} in \ref{sec:supportMaterial}) show that the flow accelerates in the outer layer due to the favorable pressure gradient. However, as shown in the present results and discussed by \citet{spina1994physics}, in supersonic boundary layers, the density decreases faster than the velocity increases. Hence, the momentum profiles decrease downstream for the S-U-F region.

\begin{figure}[H]
	\centering
	\includegraphics[trim={0.0cm 0cm 0cm 0cm},clip,width=0.99\textwidth]{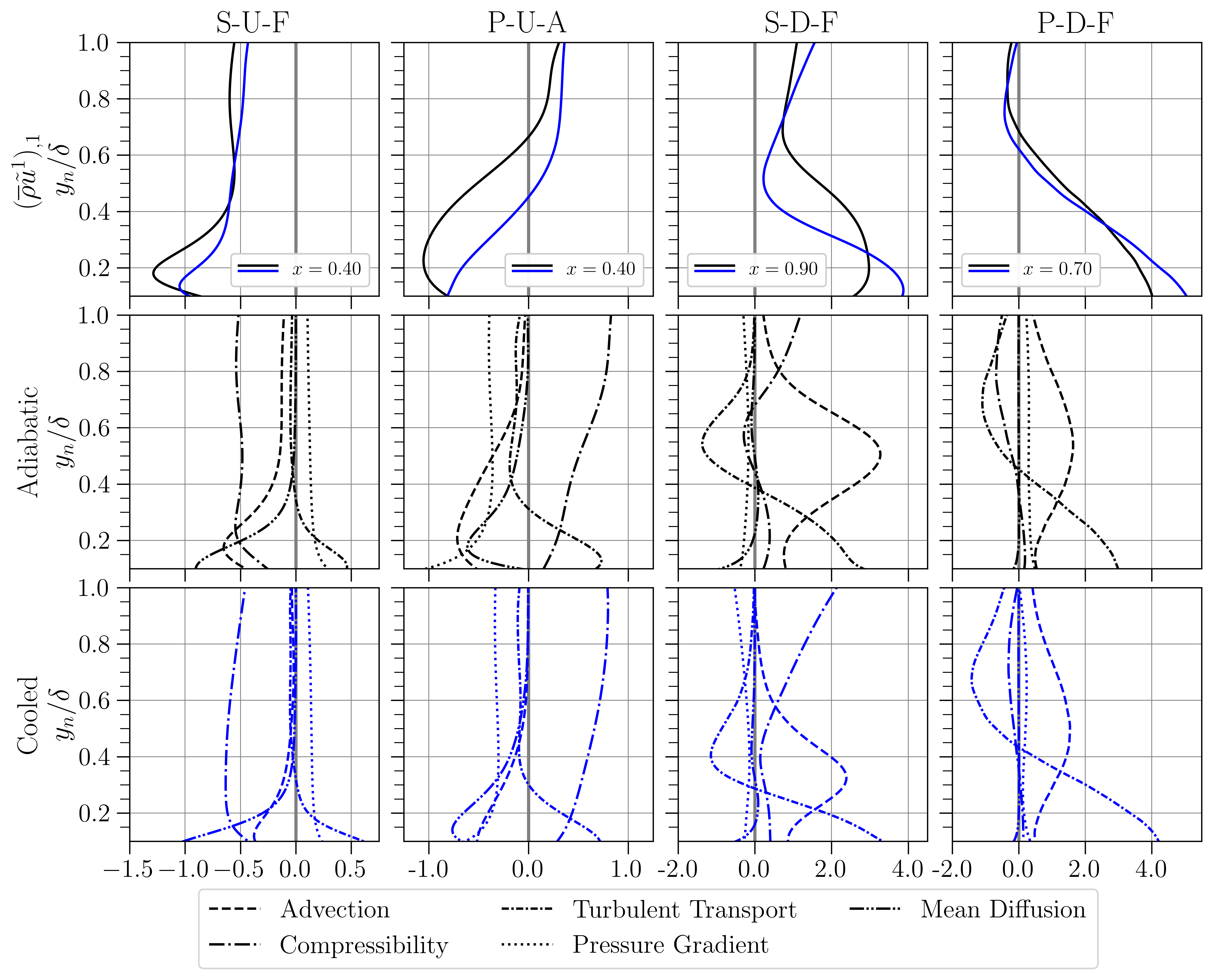}
	\caption{Balance of streamwise momentum at different locations along the blade. The first row shows the left hand side of equation \ref{eq:momentum_balance}, whereas the second and third row present their right hand sides for the adiabatic and isothermal cases, respectively. Each column depicts the individual terms of particular regions along the boundary layer (S-U-F, P-U-A, S-D-F, P-D-F).}
	\label{fig:momentum_balance}
\end{figure}

A complementary discussion can be made through Fig.\ \ref{fig:momentum_balance} that shows the balance of streamwise momentum in the boundary layer for different blade positions. This analysis allows a physical interpretation of the different mechanisms responsible for the changes in streamwise momentum (Eq.\ \ref{eq:momentum_equation} for $i=1$). In the present analysis, a Favre decomposition is applied only to convective terms while a Reynolds decomposition is employed for the remaining terms. This strategy has been used by \cite{huang1995compressible} and  \cite{lele2021turbulence} to enhance the interpretation of individual terms in the balance equations. Also, the viscosity fluctuation is neglected since it becomes relevant only at higher Mach numbers. The streamwise derivative of the streamwise momentum $(\overline{\rho} \tilde{u}^1)_{,1}$ is isolated and the balance equation is written as
\begin{equation}\label{eq:momentum_balance}
(\overline{\rho}\tilde{u}^1)_{,1} = \underbrace{-\dfrac{\tilde{u}^2(\overline{\rho}\tilde{u}^1)_{,2}}{\tilde{u}^1}}_\text{Advection}\, \underbrace{-\overline{\rho}(\tilde{u}^j)_{,j}}_\text{Compress.}\, \underbrace{-\dfrac{(\overline{\rho u''^{1}u''^{j}})_{,j}}{\tilde{u}^1}}_\text{Turb. Transport} \,\underbrace{-\dfrac{(g^{1j}\overline{p})_{,j}}{\tilde{u}^1}}_\text{Press. Grad.}  +\underbrace{\dfrac{(\overline{\tau^{1j}})_{,j}}{\tilde{u}^1}}_\text{Mean Diff.} \mbox{ .}
\end{equation} 

To enable the balance analysis in local tangential and wall-normal directions, the contravariant formulation is employed to account for the variation of the basis vectors. The terms on the right hand side of Eq. \ref{eq:momentum_balance}, from left to right, stand for the  wall-normal advection, compressibility, turbulent transport, pressure gradient, and mean diffusion, respectively. Here, the advection term represents the impact of wall-normal transport of the streamwise momentum. The first row of Fig.\ \ref{fig:momentum_balance} contains the left hand side of Eq. \ref{eq:momentum_balance}, whereas the second and third rows contain the right hand sides for the adiabatic and isothermal cases, respectively. Each column depicts the individual terms of different regions along the boundary layer (S-U-F, P-U-A, S-D-F, P-D-F). 
 
In the S-U-F region (1st column of Fig.\ \ref{fig:momentum_balance}), one can see that the pressure gradient and the turbulent transport terms display positive values, leading to an increase in streamwise momentum. While the former term acts along the entire boundary layer, the latter is relevant only near the wall. The mean diffusion and advection terms cause a reduction in momentum mostly near the wall. However, the compressibility term acts in the same fashion, but along the entire boundary layer. This term is related to the divergence of velocity $(\tilde{u}^j)_{,j}$, being an important component of the momentum balance for supersonic boundary layers affected by streamwise pressure gradients, as discussed by \cite{spina1994physics}. In the present favorable-pressure-gradient region, the data indicate a bulk dilatation effect $(\tilde{u}^j)_{,j}>0$ which leads to a reduction in the streamwise momentum due to the negative signal in Eq. \ref{eq:momentum_balance}.

On the pressure side, upstream of the SBLI (P-U-A), Fig.\ \ref{fig:momentumx} shows that the streamwise momentum profiles decrease near the wall (black arrows) and increase away from the wall (blue arrows). This counterintuitive trend occurs for both the adiabatic and isothermal cases. This behavior can be explained by Eq.\ \ref{eq:momentum_balance} and Fig.\ \ref{fig:momentum_balance}. Data presented in the column for the P-U-A region show that the streamwise derivative of the streamwise momentum $(\overline{\rho} \tilde{u}^1)_{,1}$ is negative near the wall and positive away from it. The mean diffusion and advection terms are negative near the wall and become negligible in the outer layer. On the other hand, the turbulent transport is positive near the wall, being also negligible away from it. The adverse pressure gradient contributes to the reduction of momentum along the entire boundary layer, but compressibility leads to a gain in momentum, especially in the outer layer, where this term is more pronounced than the pressure gradient. The balance of all these terms leads to the behavior observed in Fig.\ \ref{fig:momentumx}. 

Downstream of the SBLIs, the boundary layers develop under favorable pressure gradient on both sides of the blade. In Fig.\ \ref{fig:momentumx}, the momentum profiles for these two regions (S-D-F and P-D-F) increase downstream, with the pressure side (P-D-F) profiles being fuller than those of the suction side (S-D-F). The individual terms of the momentum balance in Fig.\ \ref{fig:momentum_balance} show that downstream the SBLI, both the mean diffusion and pressure gradient terms are negligible compared to advection, compressibility and turbulent transport. The advection term leads to an increase in streamwise momentum along the entire boundary layer. However, turbulent transport increases momentum near the wall and reduces it away from it. The compressibility term is relevant only in the outer layer and leads to bulk compression and dilatation on the suction and pressure sides, respectively. This difference in compressibility terms occurs because, on the suction side, the pressure gradient is adverse at $x=0.9$, only becoming favorable downstream, as shown in Fig.\ \ref{fig:integral_parameters}. In summary, the increase in streamwise momentum is not caused by the favorable pressure gradients downstream of the SBLIs. Instead, advection and turbulent transport are the primary mechanisms responsible for the momentum increase in the boundary layers in the S-D-F and P-D-F regions. The relevance of these individual mechanisms is discussed in the following section.

\subsubsection{Physical mechanisms of supersonic boundary layers under curved surfaces}

\begin{figure}[H]
	\centering
	\includegraphics[trim={0.0cm 0cm 0cm 0cm},clip,width=0.99\textwidth]{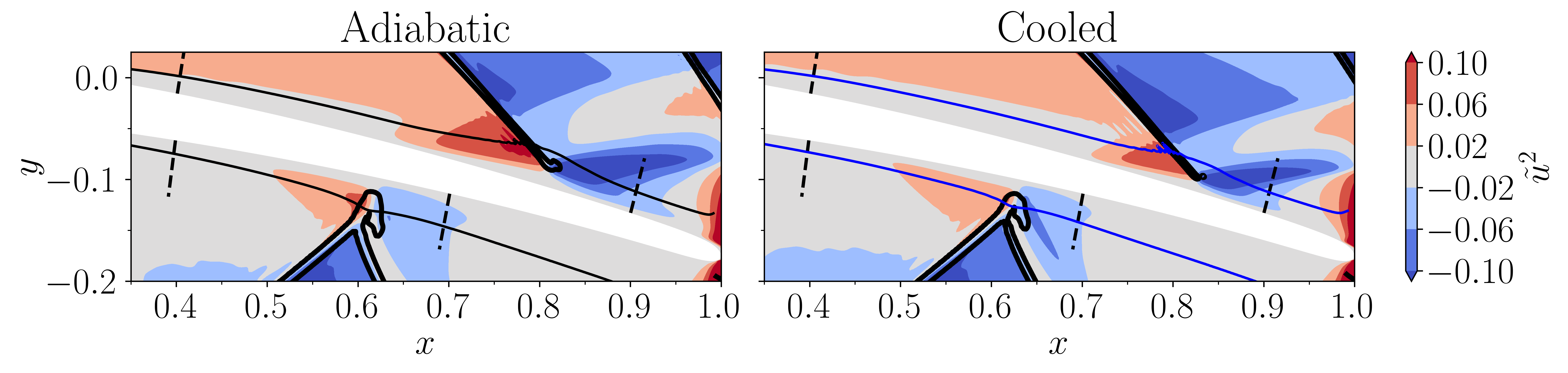}
	\caption{Mean wall-normal velocity contours along the adiabatic and isothermal blades. The solid black and blue lines on each plot represent the boundary layer thickness of the respective cases. The dashed black lines represent the positions where the momentum balance of Fig.\ \ref{fig:momentum_balance} is evaluated, and the thicker black lines depict isolines of $|\nabla\rho|$ to allow the visualization of the shock-waves.}
	\label{fig:un_mean}
\end{figure}

The advection, more specifically the wall-normal transport of streamwise momentum, is an important mechanism that contributes to changes in streamwise momentum for supersonic boundary layers developing over curved surfaces. As shown in Eq.\ \ref{eq:momentum_balance}, this term is proportional to the wall-normal velocity $\tilde{u}^2$ and the wall-normal derivative of the streamwise momentum $(\overline{\rho}\tilde{u}^1)_{,2}$. As can be seen in Fig.\ \ref{fig:momentumx}, the latter term is positive for all boundary layers analyzed here. However, the former term changes the sign downstream of the SBLIs, as shown in Fig.\ \ref{fig:un_mean}. This figure shows that the wall-normal velocity is positive everywhere upstream of the SBLIs, but becomes negative after the incident shocks.  This change in the wall-normal velocity, caused by the redirection of flow towards the wall, alters the advection mechanism and leads to an increase in streamwise momentum downstream of the SBLIs.
\begin{figure}[H]
	\centering
	\includegraphics[trim={0.0cm 0cm 0cm 0cm},clip,width=0.99\textwidth]{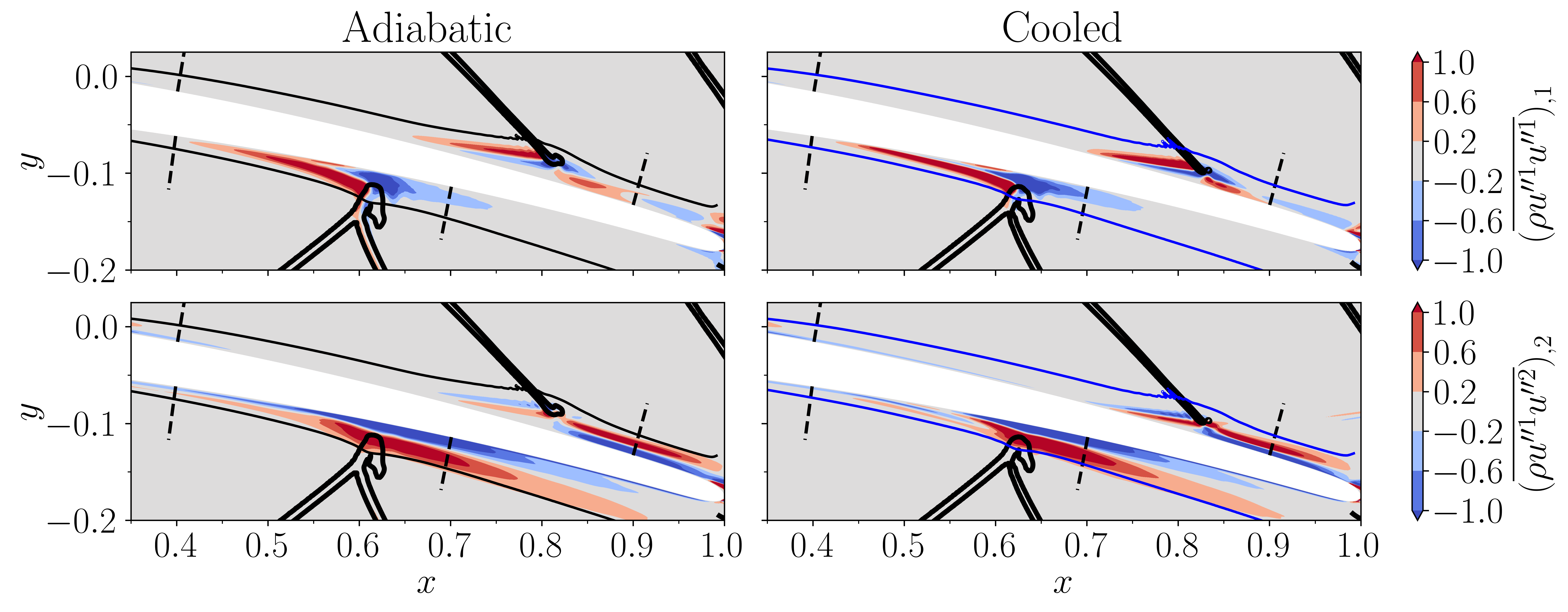}
	\caption{Derivative of the Reynolds stresses contours along the adiabatic and isothermal blades. The streamwise component is represented at the top, whereas the shear component is at the bottom. See Fig.\ \ref{fig:un_mean} for details of the auxiliary wall-normal lines.}
	\label{fig:drhouiujdxj}
\end{figure}

Turbulent transport is another key mechanism in supersonic turbulent boundary layers, particularly downstream of SBLIs, as observed in Fig.\ \ref{fig:momentum_balance}. In the S-D-F and P-D-F regions, turbulent transport accelerates the flow in the inner layer and decelerates it in the outer layer. This phenomenon can be explained by the shear layers formed downstream of the SBLIs. From Eq.\ \ref{eq:momentum_balance}, the turbulent transport depends on the tangential velocity $\tilde{u}^1$, shown in Fig.\ \ref{fig:ut}, and on the sum of the derivatives of Reynolds stresses  $(\overline{\rho u''^{1}u''^{j}})_{,j}$. This latter term can be separated into streamwise and shear components, as shown in the top and bottom rows of Fig.\ \ref{fig:drhouiujdxj}, respectively. The sign of the tangential velocity is positive throughout the boundary layer thickness, indicating that it is not responsible for the change in behavior observed in the turbulent transport. Both terms of the Reynolds stress derivatives are impacted by the SBLIs, with the shear term dominating over the streamwise term downstream of the SBLI (Fig.\ \ref{fig:turb_transp_mean}). Velocity profiles with inflection points, such as those of Fig.\ \ref{fig:ut}, can display shear layers embedded within the boundary layers \citep{priebe2012,fang2020turbulence,leandro2024}. Downstream of the SBLIs, the derivative of the Reynolds shear stress is positive along the shear layer, while negative values are observed in the near-wall boundary layer region. These results confirm the flow acceleration near the wall, as expected, since the boundary layers are subjected to a favorable pressure gradient. However, the free shear layers decelerate the flow away from the wall, leading to a change in sign observed in the turbulent transport mechanism downstream of the SBLIs in Fig.\ \ref{fig:momentum_balance}.

Finally, as observed in Fig.\ \ref{fig:momentum_balance}, the compressibility term is also a prominent component affecting the streamwise momentum. This term dominates the momentum balance (Eq.\ \ref{eq:momentum_balance}) in the outer region of the boundary layers upstream of the SBLIs, regardless of the pressure gradient. This term remains relevant downstream of the SBLIs as well. The compressibility term is composed of the mean density and the divergence of velocity as shown in Eq.\ \ref{eq:momentum_balance}. Through the continuity equation, it can be expressed as a function of the density gradient as:
\begin{equation}\label{eq:continuity_divu}
\overline{\rho}(\tilde{u}^j)_{,j} = -\tilde{u}^1(\overline{\rho})_{,1} -\tilde{u}^2(\overline{\rho})_{,2} \mbox{ .}
\end{equation}

\begin{figure}[H]
	\centering
	\includegraphics[trim={0.0cm 0cm 0cm 0cm},clip,width=0.99\textwidth]{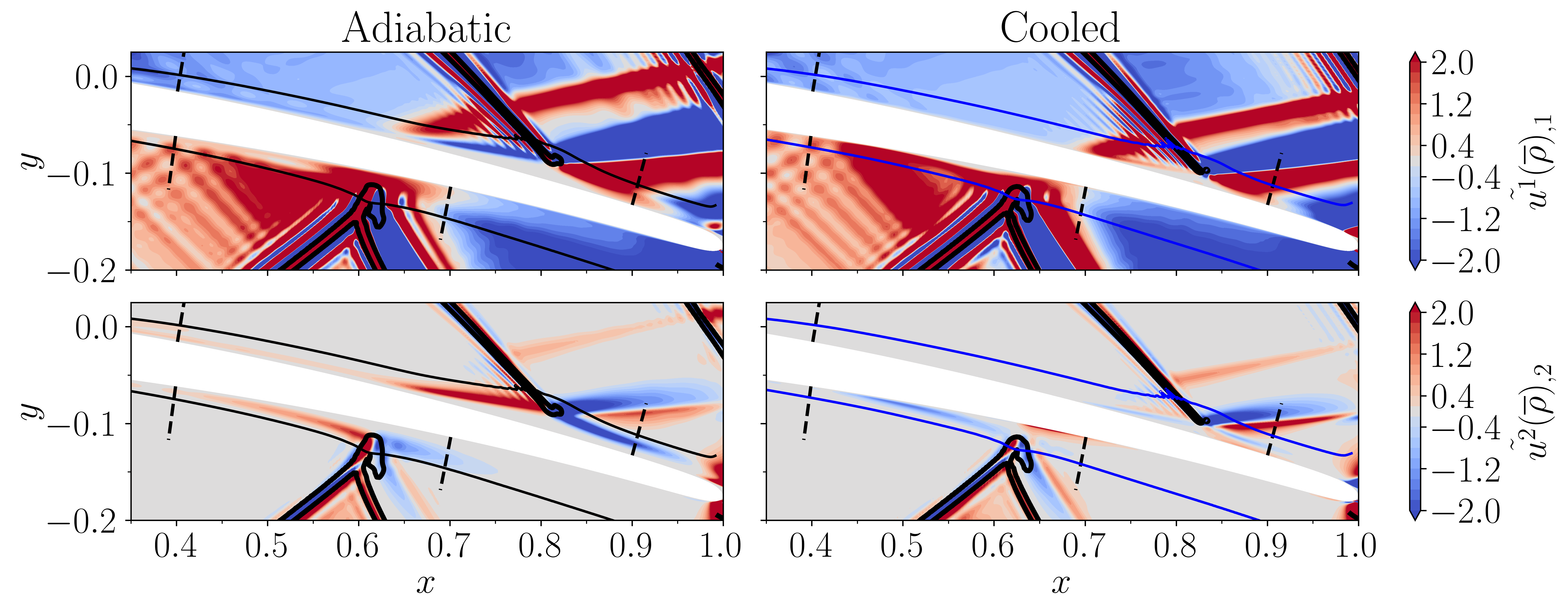}
	\caption{Streamwise (top) and wall-normal (bottom) terms of the compressibility term (Eq.\ \ref{eq:continuity_divu}) for adiabatic and isothermal blades. See Fig.\ \ref{fig:un_mean} for details of the auxiliary wall-normal lines.}
	\label{fig:uidrhodxj}
\end{figure}

In the above equation, the streamwise velocity $\tilde{u}^1$ is positive along the boundary layers away from the recirculation regions. However, the wall-normal component $\tilde{u}^2$ changes direction downstream of the shock waves and points toward the surface, as shown in Fig.\ \ref{fig:un_mean}. The compressibility also depends on the density gradient, which can be assessed by inspection of Fig.\ \ref{fig:density}. Upstream of the SBLIs, this figure shows that the streamwise derivative of density is negative on the suction side and positive on the pressure side due to the flow expansion and compression, respectively. This behavior is observed for both adiabatic and isothermal blades, and it results from the pressure gradients induced by the convex and concave curvatures on the suction and pressure sides, respectively. Hence, upstream of the SBLI, the first term of Eq.\ \ref{eq:continuity_divu} is negative on the suction side and positive on the pressure side, as shown in Fig.\ \ref{fig:uidrhodxj}. This figure presents the contours of the individual terms that comprise the compressibility term. Immediately downstream of the SBLIs, a series of compressions and expansions directly impact the compressibility of the boundary layers due to the shock waves. Figure \ref{fig:uidrhodxj} also shows that the first term of Eq.\ \ref{eq:continuity_divu} plays a dominant role in compressibility, even though the wall-normal derivative of density is higher than its streamwise component. This occurs because the wall-normal velocity is considerably lower than the streamwise component. The full compressibility term is shown in \ref{sec:supportMaterial} as Fig.\ \ref{fig:divU_mean} for completeness, being obtained by the sum of both components.

\section{Conclusions}\label{sec:conclusions}

The present work investigates supersonic boundary layers developing over the curved surfaces of a turbine blade. The effects of pressure gradient and SBLIs on the boundary layers are assessed for adiabatic and isothermal (cooled) surfaces. Results are obtained by wall-resolved LES performed for a linear turbine cascade with inlet Mach and Reynolds numbers of $M_\infty = 2.0$ and  $Re_\infty = 200,000$, respectively. The boundary layers are tripped near the airfoil leading edge to induce bypass transition, and the wall-to-inlet temperature ratio of the isothermal case is $T_w/T_\infty  = 0.75$. 

The thermal boundary conditions exhibit an important role on the density and temperature distributions along the boundary layers. Surface cooling leads to higher densities and lower temperatures near the wall, which also reduces viscosity. Opposite trends are observed for the adiabatic case as a result of heat transfer to the gas instead of the blade, due to aerodynamic heating. Hence, the isothermal case has a higher local Reynolds number at the wall and fuller streamwise momentum profiles, which are less prone to separation, as observed in the chordwise distribution of the shape factor. In the same fashion, the SBLI-induced separation bubble on the suction side is also smaller and thinner for the isothermal case compared to the adiabatic one.

The curved walls of the turbine blade induce favorable and adverse pressure gradients on the suction (convex) and pressure (concave) walls, upstream of the SBLIs. The spatial variation of the density profiles computed along the turbulent boundary layers shows that the favorable pressure gradient leads to a gas expansion, while the adverse pressure gradient causes its compression. In general, FPGs accelerate the flow along the boundary layers while APGs decelerate. However, a reduction in streamwise momentum is observed on the suction side, upstream of the SBLI, despite the favorable pressure gradient. This occurs because, in the supersonic portion of the boundary layer, the density decreases faster than the velocity increases. Moreover, an analysis of the streamwise momentum balance equation shows that compressibility effects, which arise due to the pressure gradients, have a leading impact in the momentum distribution along the boundary layer. In this case, the favorable pressure gradient induces a bulk dilation while the adverse pressure gradient causes a bulk compression. The former mechanism reduces the streamwise momentum, while the latter increases.

Downstream of the SBLIs, the boundary layers develop under favorable pressure gradient, and momentum profiles show a flow acceleration on both suction and pressure sides for the adiabatic and isothermal cases. However, the leading mechanism responsible for the gain in momentum is not the favorable pressure gradient, but the turbulent transport and the wall-normal mean advection. The turbulent transport appears near the wall due to boundary layers and free shear layers downstream of the separation bubbles, on both sides of the airfoil. On the other hand, the wall-normal advection is significantly impacted by the reflected shock waves that redirect the mean wall-normal velocity toward the blade. The SBLIs also play an important role in the compressibility term through a series of compressions and expansions over the separation bubble.

\section*{Acknowledgments}

The authors acknowledge Fundação de Amparo à Pesquisa do Estado de São Paulo, FAPESP, for supporting this work under grants No. 2013/08293-7, 2019/17874-0, 2021/06448-0, 2022/00464-6, 2024/04341-1 and 2024/21935-2, and the Air Force Office of Scientific Research, AFOSR, for supporting this work under grant FA9550-23-1-0615. Access to the HPC resources of IDRIS, TGCC, and CINES under the allocation A0172A12067 made by GENCI are also acknowledged.

\appendix
\section{Tensor calculus and generalized curvilinear coordinates}\label{sec:appendixATensor}

In the present work, the flow governing equations are written in generalized curvilinear coordinates ($\xi^j, j = 1,2,3$). In this form, the variation of the basis vectors is accounted for in the covariant derivative \citep{aris1990vectors}, which defines a derivative along the tangent vectors of a manifold. This is obtained by applying the product rule to a vector represented by contravariant components, as follows:
\begin{equation}\label{eq:derivative_vector}
\begin{aligned}
    \partial_j\mathbf{A} &= \partial_j(A^i\mathbf{E}_i)\\
    &= \mathbf{E}_i\partial_jA^i + A^i\partial_j\mathbf{E}_i \quad \text{(product rule)}\\
    &= \mathbf{E}_i\partial_jA^i + A^i\Gamma_{ij}^k\mathbf{E}_k\quad\text{ (where } \partial_j\mathbf{E}_i = \Gamma_{ij}^k\mathbf{E}_k\text{)} \\
    &= \mathbf{E}_i\partial_jA^i + A^k\Gamma_{kj}^i\mathbf{E}_i\quad\text{(relabeling indices $i$ and $k$)} \\
    &= (\partial_jA^i + A^k\Gamma_{kj}^i)\mathbf{E}_i \\
    &= (A^i)_{,j}\mathbf{E}_i \mbox{ .}
\end{aligned}
\end{equation}
Here, $A^i$ is the contravariant component of vector $\mathbf{A}$, $\mathbf{E}_i = \pdv{\mathbf{x}}{\xi^i}$ is the covariant basis vector, $\mathbf{x} = (x^1,x^2,x^3) = (x,y,z)$ is the Cartesian basis vector and $\Gamma_{kj}^i$ is the Christoffel symbol of the second kind, defined as:
\begin{equation}
    \Gamma_{jk}^i = \dfrac{g^{il}}{2}(\partial_kg_{jl} + \partial_jg_{kl} -\partial_lg_{jk}) \mbox{ .}
\end{equation}
The covariant and contravariant metric tensors are given by $g_{ij} = \pdv{\xi^i}{x^k}\pdv{\xi^j}{x^k}$ and $g^{ij} = \pdv{x^k}{\xi^i}\pdv{x^k}{\xi^j}$, respectively. As shown by \citet{aris1990vectors}, the covariant derivative of a contravariant tensor of arbitrary order can be written as  
\begin{equation}
A^{ij\dots k}_{,q} = \partial_qA^{ij\dots k} +\Gamma_{aq}^iA^{aj\dots k} + \Gamma_{aq}^jA^{ia\dots k} +\cdots +\Gamma_{aq}^kA^{ij\dots a} \mbox{ .}
\end{equation}
The flow governing equations are obtained by writing their components in contravariant form along with their respective basis vectors. This is followed by the evaluation of the covariant derivatives, as shown in Eq. \ref{eq:derivative_vector}. Below, the covariant derivatives of first- and second-order tensors are presented for the mass and momentum fluxes, as they appear in Eqs. \ref{eq:continuity_equation} and \ref{eq:momentum_equation}
\begin{equation}
    (\rho u^j)_{,j} = \partial_j(\rho u^j) +\Gamma_{kj}^j\rho u^k \mbox{ ,}
\end{equation}
\begin{equation}
    (\rho u^iu^j)_{,j} = \partial_j(\rho u^iu^j) +\Gamma_{kj}^i\rho u^ku^j +\Gamma_{kj}^j\rho u^iu^k \mbox{ .}
\end{equation}

\section{Analysis of spanwise domain size and grid resolution}\label{sec:appendixWallUnits}

\begin{figure}[H]
    \centering
    \includegraphics[trim={0.0cm 0cm 0cm 0cm},clip,width=0.99\textwidth]{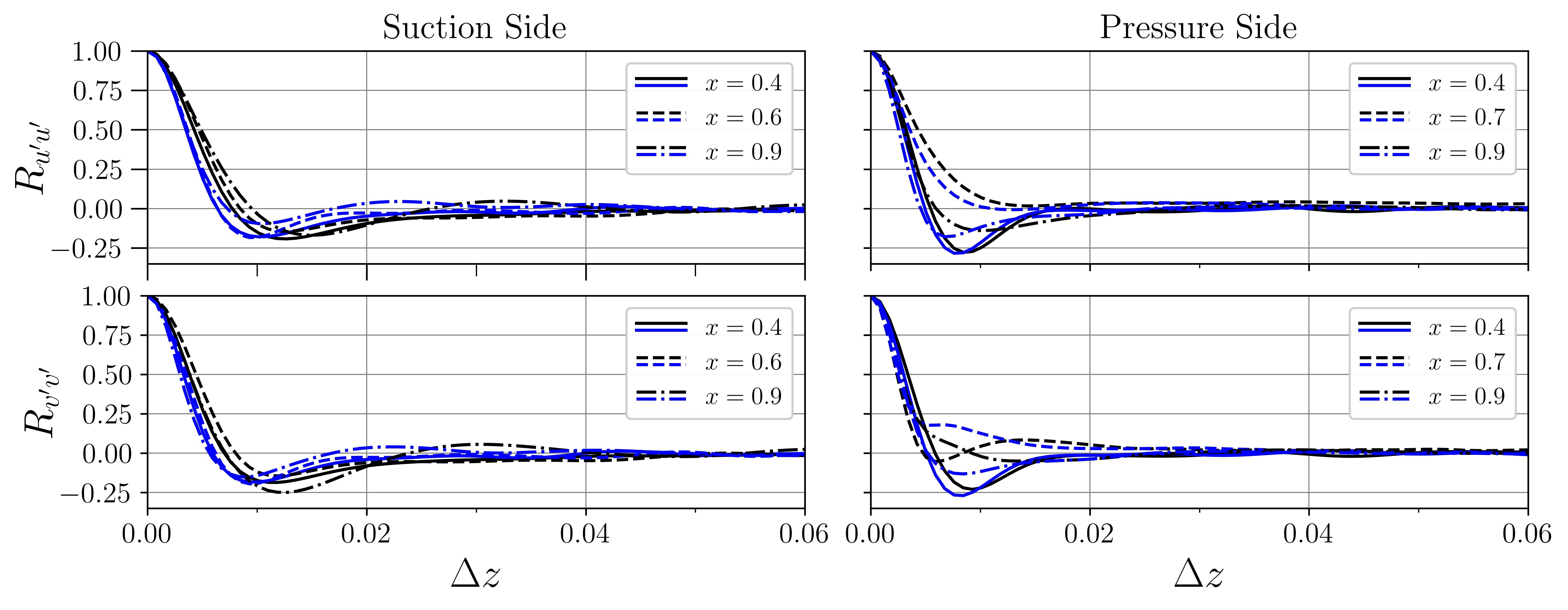}
    \caption{Two-point spanwise correlations of horizontal (top) and vertical (bottom) velocity fluctuations at different chord locations on the suction side (left) and pressure side (right) for adiabatic (black) and cooled (blue) cases.}
    \label{fig:spanwise_correlation}
\end{figure}

Figure \ref{fig:spanwise_correlation} shows the two-point spanwise correlation of horizontal ($u'$) and vertical ($v'$) velocity fluctuations at $y^+ \approx 5$ for different chord locations, both upstream and downstream of the separation bubbles, on the suction and pressure sides. Results are presented for the horizontal and vertical velocity components in the top and bottom plots, respectively. The correlations exhibit a rapid decay along the span on both sides of the airfoil, demonstrating that the spanwise domain is sufficiently wide for the development of turbulence dynamics.

\begin{figure}[H]
    \centering
    \includegraphics[trim={0.0cm 0cm 0cm 0cm},clip,width=0.99\textwidth]{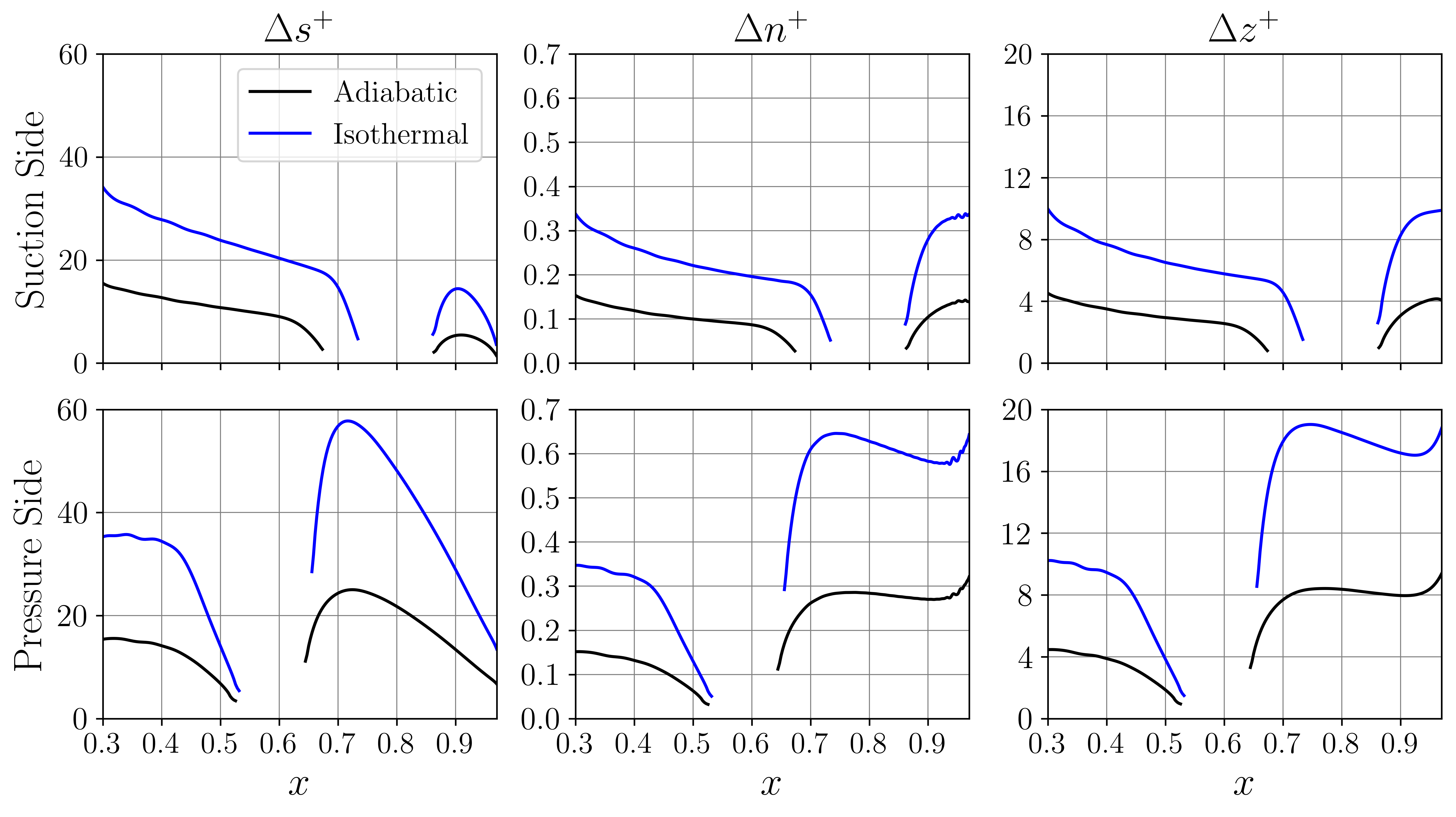}
    \caption{Chordwise distribution of grid spacing in terms of wall units in streamwise (left), wall-normal (middle) and spanwise (right) directions on the suction (top) and pressure sides (bottom), for adiabatic (black) and cooled (blue) blades.}
    \label{fig:delta_plus}
\end{figure}

To evaluate the grid resolution, the chordwise distribution of grid spacing is shown in Fig.\ \ref{fig:delta_plus} in terms of wall units. The values of $\Delta s^+$, $\Delta n^+$ and $\Delta z^+$ indicate adequate near-wall mesh resolution, satisfying the guidelines for wall-resolved LES \citep{georgiadis2010large}. 

\begin{figure}[H]
    \centering
    \includegraphics[trim={0.0cm 0cm 0cm 0cm},clip,width=0.99\textwidth]{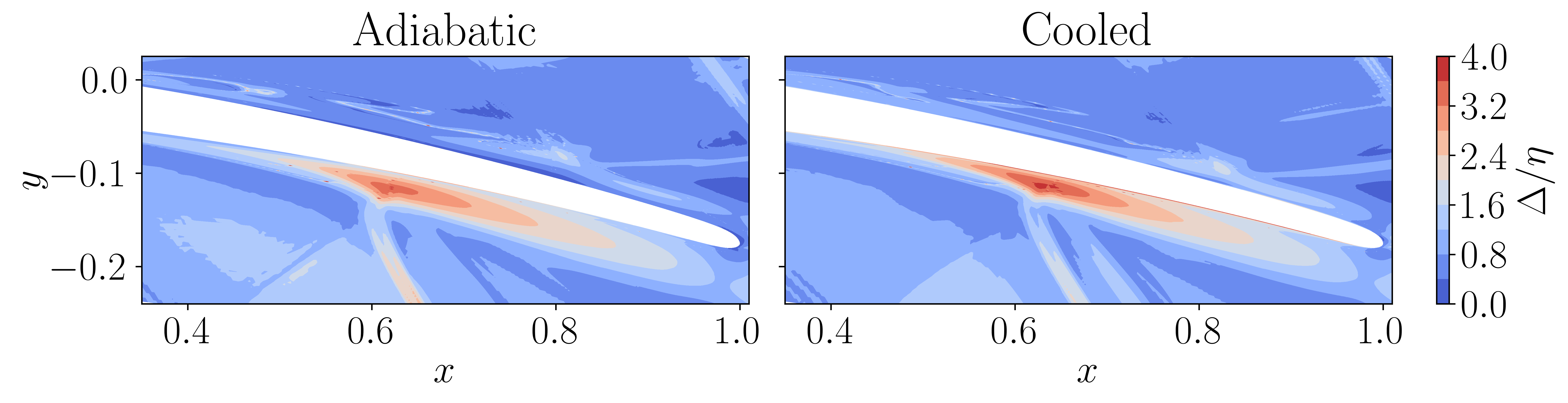}
    \caption{Ratio between grid spacing (cubic root of cell volume) and  estimated Kolmogorov length scale $\eta$ for adiabatic (left) and cooled (right) blades.}
    \label{fig:kolmogorov}
\end{figure}

To further evaluate the grid quality, the ratio of grid spacing ($\Delta$), calculated as the cubic root of the cell volume, to the estimated Kolmogorov length scale ($\eta$) is presented in Fig.\ \ref{fig:kolmogorov}. The latter is estimated as $\eta = (\overline{\nu}^3/\epsilon)^{1/4}$, where $\overline{\nu} = \overline{\mu}/\overline{\rho}$ is the kinematic viscosity and $\epsilon = \overline{\tau'^{ij}(u'^i)_{,j}}$ is the turbulent kinetic energy dissipation rate. The maximum computed values are around 4, occurring along and downstream of the SBLI on the pressure side. These results indicate that the present grid has adequate resolution for a wall-resolved LES, in agreement with other studies (\cite{schiavo2015large,lui2024mach}).

\section{Supporting data}\label{sec:supportMaterial}

\begin{figure}[H]
	\centering
	\includegraphics[trim={0.0cm 0cm 0cm 0cm},clip,width=0.99\textwidth]{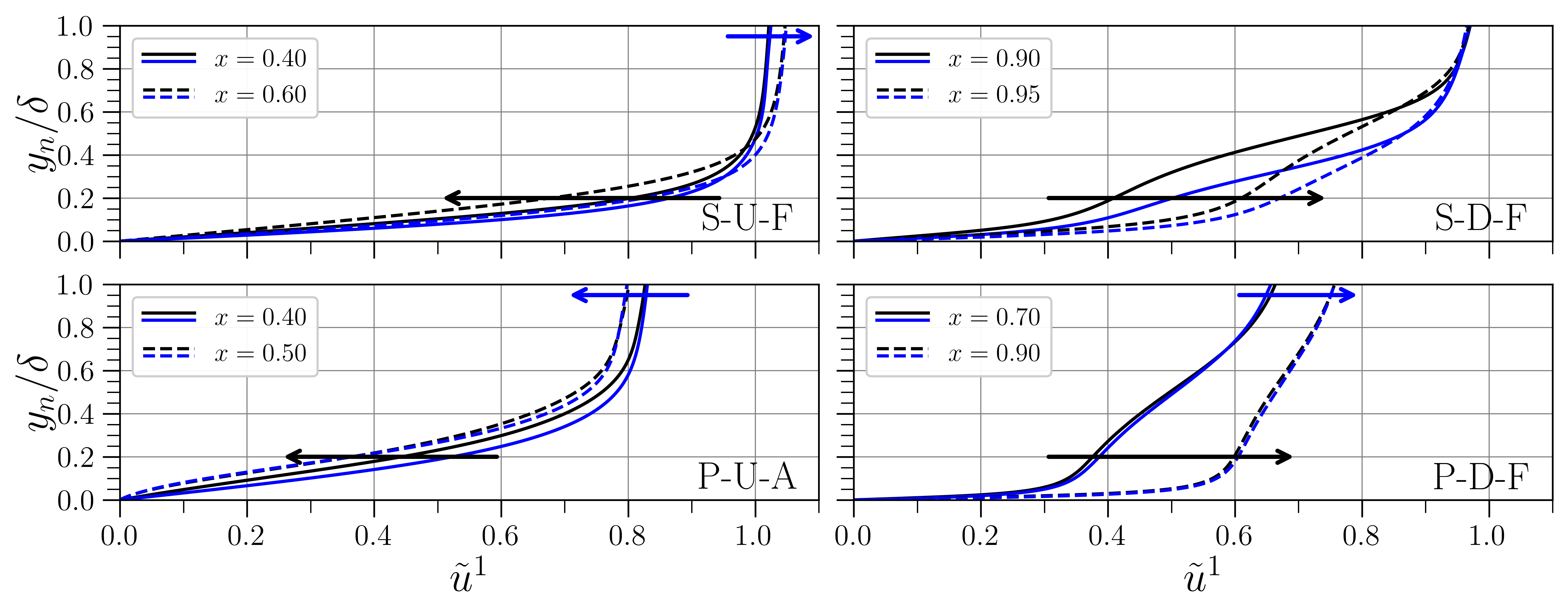}
	\caption{Streamwise (tangential) velocity profiles for adiabatic (black lines) and isothermal (blue lines) cases for the locations specified in Fig.\ \ref{fig:bl_thickness}. From left to right, the four locations are S-U-F, P-U-A, S-D-F, P-D-F. The solid/dashed lines represent the upstream/downstream positions, respectively.}
	\label{fig:ut}
\end{figure}

\begin{figure}[H]
	\centering
	\includegraphics[trim={0.0cm 0cm 0cm 0cm},clip,width=0.99\textwidth]{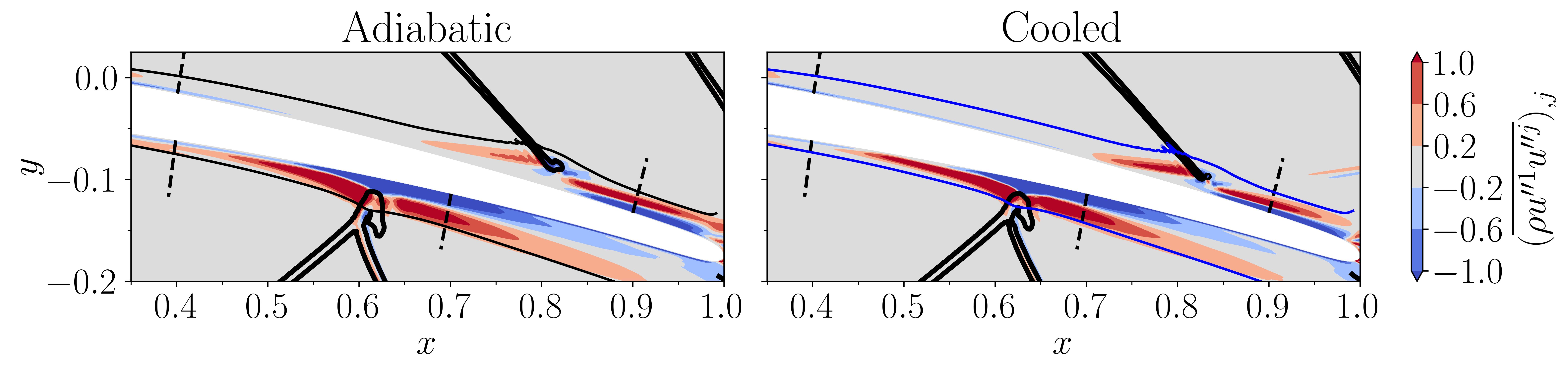}
	\caption{Contours of the summation of Reynolds stresses derivatives along the adiabatic and isothermal blades. The solid black and blue lines on each plot represent the boundary layer thickness of the respective cases. The dashed black lines represent the positions where the momentum balance of Fig.\ \ref{fig:momentum_balance} is evaluated, and the thicker black lines depict isolines of $|\nabla\rho|$ to allow the visualization of the shock-waves.}
	\label{fig:turb_transp_mean}
\end{figure}

\begin{figure}[H]
	\centering
	\includegraphics[trim={0.0cm 0cm 0cm 0cm},clip,width=0.99\textwidth]{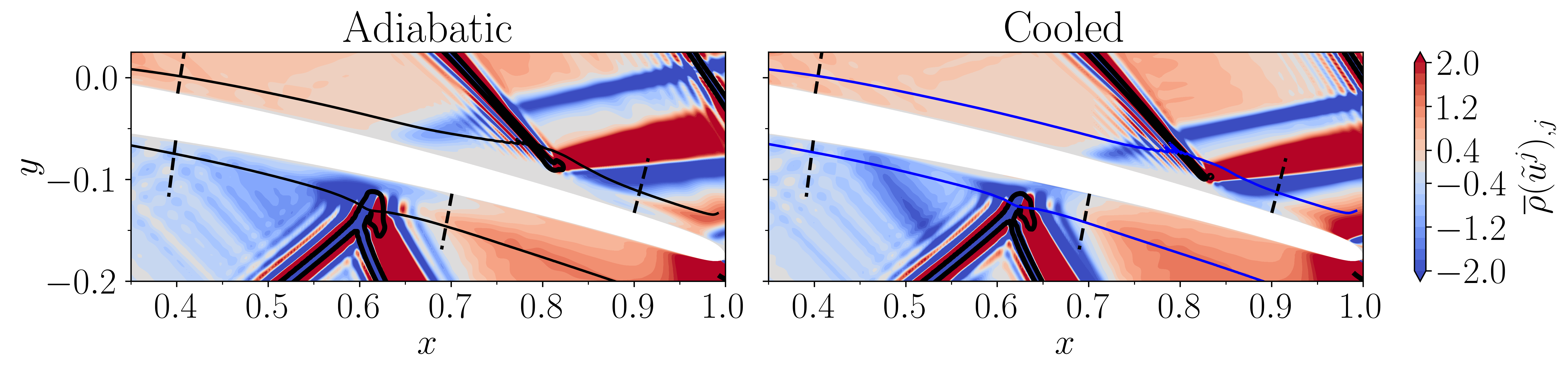}
	\caption{Contours of the compressibility term along the adiabatic and isothermal blades. See Fig.\ \ref{fig:turb_transp_mean} for details of the auxiliary lines.}
	\label{fig:divU_mean}
\end{figure}

The tangential velocity profiles (Fig.\ \ref{fig:ut}) are presented here as supporting material for the discussion of the results. Similarly, iso-contours of the summation of the Reynolds stress derivatives (Fig.\ \ref{fig:turb_transp_mean}) and the full compressibility term (Fig.\ \ref{fig:divU_mean}) are also provided to assist in the discussion.

\bibliographystyle{elsarticle-harv}
\bibliography{refs}

\end{document}